\documentclass[final,3p, twocolumn]{elsarticle}

\usepackage{hyperref}
\usepackage{empheq}
\usepackage{multirow}
\journal{Computer Physics Communications}

\usepackage{xcolor}

\newcommand{\new}[1]{}

\newcommand{\PHARE}{\texttt{PHARE}}
\DeclareUnicodeCharacter{0301}{\'{e}}
\begin{document}

\begin{frontmatter}

\title{PHARE : Parallel hybrid particle-in-cell code with patch-based adaptive mesh refinement}


\author[lpp]{Nicolas Aunai\corref{mycorrespondingauthor}}
\cortext[mycorrespondingauthor]{Corresponding author}
\ead{nicolas.aunai@lpp.polytechnique.fr}

\author[lpp]{Roch Smets}
\author[lerma]{Andrea Ciardi}
\author[lpp]{Philip Deegan}
\author[lpp]{Alexis Jeandet}
\author[lis]{Thibault Payet}
\author[lpp]{Nathan Guyot}
\author[lpp]{Loic Darrieumerlou}

\address[lpp]{Laboratoire de Physique des Plasmas (LPP), CNRS, Observatoire de Paris, Sorbonne Université, Université Paris-Saclay, École polytechnique, Institut Polytechnique de Paris, 91120 Palaiseau, France}

\address[lerma]{Sorbonne Université, Observatoire de Paris, Université́ PSL, CNRS, LERMA, F-75005 Paris, France}

\address[lis]{Aix Marseille Univ, CNRS, LIS, Marseille, France}

%\author{Nicolas Aunai}
%\author{Roch Smets}
%\author{Philip Deegan}
%\author{Alexis Jeandet}
%\author{Nathan Guyot}
%\author{Loic Darrieumerlou}
%\address{Laboratoire de Physique des Plasmas (LPP), CNRS, Observatoire de Paris, Sorbonne %Université, Université Paris-Saclay, École polytechnique, Institut Polytechnique de Paris, %91120 Palaiseau, France}
%%\fntext[myfootnote]{Since 1880.}
%
%
%\author{Andrea Ciardi}
%\address{Sorbonne Université, Observatoire de Paris, Université́ PSL, CNRS, LERMA, F-75005 %Paris, France}
%
%
%
%\author{Thibault Payet}
%\address{Aix Marseille Univ, CNRS, LIS, Marseille, France}
%




\begin{abstract}
Modeling multi-scale collisionless magnetized processes constitutes an important numerical challenge.
By treating electrons as a fluid and ions kinetically, the so-called hybrid Particle-In-Cell (PIC) codes represent a promising intermediary between fully kinetic codes, limited to model small scales and short durations, and magnetohydrodynamic codes used large scale.
However, simulating processes at scales significantly larger than typical ion particle dynamics while resolving sub-ion dissipative current sheets remain extremely difficult.
This paper presents a new hybrid PIC code with patch-based adaptive mesh refinement.
Here, hybrid PIC equations are solved on a hierarchy of an arbitrary number of  Cartesian meshes of incrementally finer resolution dynamically mapping regions of interest, and with a refined time stepping.
This paper presents how the hybrid PIC algorithm is adapted to evolve such mesh hierarchy and the validation of the code on a uniform mesh, fixed refined mesh and dynamically refined mesh.
\end{abstract}

%\begin{keyword}
%\texttt{elsarticle.cls}\sep \LaTeX\sep Elsevier \sep template
%\MSC[2010] 00-01\sep  99-00
%\end{keyword}

\end{frontmatter}

%\linenumbers

\section{Introduction}
The numerical modeling of magnetized space and laboratory plasmas represents an important computational challenge that is generally linked to the need to capture the multi-scale nature of plasma dynamics. While kinetic equations such as the Vlasov-Maxwell system for collisionless plasmas can in principle describe such dynamics, from electron scales to global system scales, their use in multi-dimensions is in practice restricted to relatively small scales. For example, fully kinetic simulations of magnetic reconnection usually handle scales of a few tens of ion inertial lengths $\delta_i$, quite rarely in three dimensions, and very often with artificially reduced scale separations through modified ion to electron mass ratio and/or reduced speed of light \cite{Dargent2020,DaughtonNat2011}.
%Modeling scales very large compared to ion kinetic scales is typically done using fluid equations such as magnetohydrodynamics (MHD). 
%Many formalisms exist between fully kinetic equations and MHD.
Modeling larger scales is typically done with more approximate physical formalisms.
Among them is the hybrid formalism, in which ions are treated kinetically but electrons are modeled as a  fluid \cite{Filippychev2002, lipatov2002}. 
This allows, in principle, to transfer the computational power required to resolve electron scales into solving larger domains for longer times.

Hybrid codes have been used for modeling  large systems such as (small or artificially reduced) planetary magnetospheres.
However such large scale models typically poorly resolve the Hall scale \cite{2016JCoPh.309..295L,2017JGRA..122.2877H, guo2021, guo2020}.
Not resolving such sub-ion scale current sheets may result in errors leading to spurious macroscopic behavior \cite{2013PhPl...20d2901A}.
The main advantage such models have compared to a well resolved yet much lighter magnetohydrodynamics (MHD) counterparts is that they account for collisionless population mixing.

Modeling scales much larger than typical ion scales and in three dimensional systems yet having sub-ion scale space/time resolution remains a very challenging task for hybrid codes.
The dispersive nature of the whistler waves supported by the collisionless Ohm's law imposes very strong constraints on the time step, which scales as the square of the mesh size.
Therefore, high resolution large scale models ($\Delta \sim 0.1\delta_i$) remain computationally intensive.
In practice, such high resolution hybrid simulations have typically been done in domains that are not vastly larger than those accessible with full-PIC codes using reduced physical normalized constants. The small gain in domain size or evolution time offered by hybrid codes is thus often quickly balanced by the drawback of losing the electron kinetic physics offered by full PIC ones. In practice, hybrid codes have not been so competitive once high resolution is needed.

The hybrid equations can in practice either be solved in a Eulerian way, with so-called Vlasov hybrid codes \cite{Palmroth2018Review}, or with a Lagrangian perspective, with the so-called hybrid Particle-In-Cell (PIC) codes \cite{lapenta2012sw}.
PIC codes present the advantage over Vlasov ones to be inherently adaptive in their description of the particle distribution.
This is more efficient computationally than resolving fine structures developing for different reasons in position space and velocity space and for each of the populations composing the plasma.
Some features are also easier and/or lighter to handle in PIC than in Vlasov models, such as the possible interactions between the populations.
On the other hand, Vlasov codes present the advantage to be noise-free contrary to PIC results which are inherently based on statistical sampling of the continuous distribution function.
Recent advances have been achieved in using the Vlasov hybrid formalism to model large scales in the context of the Earth's magnetosphere \cite{Palmroth2018Review}.
Yet, simulations remain heavier than their PIC counterpart for which noise can moreover be reduced through the use of higher order interpolation kernels and increased number of macroparticles per cell. PIC codes thus constitute a promising approach for developing multi-scale models.

There has been several attempts to make hybrid PIC codes more adapted to model multi-scale systems.
Time scale disparities have been for instance addressed by decomposing the domain into different time zones, each evolving with a proper time step \cite{karimabadi2006tz}.
Evolving quantities based on local event triggers rather than at regular and global time step intervals \cite{omelchenko2006DES,Karimabadi:2007uv, omelchenkohypers2012} has also been considered.

Spatial scale disparities, on the other hand, are generally handled by adopting a non-uniform meshing.
This can be achieved either by refining the grid in the so-called adaptive mesh refinement (AMR) methods \cite{JLVay2002,JLVay2004,fujimoto2006,Fujimoto:2008jv,colella2010,muller2011,fujimoto2011,Fujimoto:2018gi}, or adapting the position of the nodes of a single mesh in the so-called moving mesh adaptation (MMA) method \cite{Lapenta:2007tj,lapenta2011democritus}.
In both cases the goal is to focus the meshing in regions developing small scale features where more computational power will be dedicated, and for which one needs to define ad-hoc detection criteria. 

Adaptive mesh refinement (AMR) has now become a mainstream feature of fluid codes \cite{fryxell2008,2012ApJS..198....7M, 2008JCoPh.227.6967T}.
It is however much less used in PIC codes.
The main reason probably stems from the algorithmic complexity in handling macroparticles across multiple grid levels. Only one AMR fully kinetic electromagnetic code is currently in use in our knowledge \cite{fujimoto2006,Fujimoto:2008jv,fujimoto2011,Fujimoto:2018gi}.

An AMR Hybrid PIC code following a hybrid block AMR method \cite{holst2007} has also been proposed \cite{muller2011}.
It is worth noticing however that the code has been mostly used with multiple grids that were fixed in time \cite{Feyerabend:2015ib,Vernisse:2017ga,Exner:2018fj,arnold2020}.

%In AMR fluid codes, the same equations are solved on all refinement levels, coarser solution being used as boundary conditions for finer levels, which solution, once obtained, is coarsened to lower levels for synchronization purposes. 
In existing AMR implementations for PIC codes, macroparticles interact only with the finest mesh at their location.
This method has several drawbacks.
The first one, shared with MMA, is that macroparticles see multiple mesh spacings as they evolve.
Either because they cross a grid boundary along their trajectory or because the region in which they currently are is being refined or coarsened in time (or streched/contracted in MMA). 
Because the compact support of the macroparticle interpolation kernel is usually constant in index-space, these macroparticles will see their physical extent change in time, potentially leading to a spurious evolution of their momentum \cite{lapenta2012sw}. 
A more important drawback of having particles exploring multiple grids is that the number of particles per cell strongly decreases in refined cells if nothing is done.
In practice this number is usually kept roughly constant by splitting macroparticles when they enter in refined regions \cite{fujimoto2006,Fujimoto:2008jv, fujimoto2011, muller2011}.
If the splitting procedure can trivially conserve the velocity distribution of macroparticles, it may not conserve their spatial distribution, and thus their contribution to moments on the mesh.
Several splitting procedures have been proposed \cite{lapenta2002rezoning,Fujimoto:2018gi,innocentiMLMD1, muller2011}.
They all dispatch refined macroparticles within some ad hoc distance from the original coarse one, usually in the same cell, randomly or not.
\new{In the current study, we employ a method recently proposed for dispatching split particles in AMR PIC codes\cite{Smets:2021dw}.
The method distributes N split particles around the original coarse macroparticle in a way that exactly (or optimally, depending on the number of split particle chosen for a given dimension and B-spline order) conserves the assignment function, and not only its values on coarse mesh nodes.
This has the advantage that the assignment function of the fine population will have the same value on the nodes uniquely found on the fine mesh, as would have the one of the coarse population at this location, at the time of the splitting.
}
%In any case, these methods do not guarantee the assignment function of the original coarse macroparticle population is equal or as close as possible to that of the refined particles. 
%Thus the contribution of the refined particles to the fine mesh will differ in some uncontrolled way from what would be that of the coarse population.
%This typically adds a significant amount of noise on the refined mesh \cite{Smets:2021dw}.
If macroparticles only interact with the finest mesh at their location, and splitting is used not to decrease their number per cell, the opposite operation, i.e. the merging of macroparticles, needs to be done as well not to stall the simulation.
Merging macroparticles is more hazardous than splitting as it generally does not preserve the local structure of the distribution function \cite{lapenta2002rezoning}, the very reason why kinetic simulations are done in the first place.

The Multi-Level-Multi-Domain (MLMD) method has later been proposed as a way to prevent previous issues \cite{innocentiMLMD1, beckMLMD2, innocentiMLMDjp, innocentiMLMDmomentum,innocentiMLMDdt}. 
Contrary to other AMR codes, the MLMD method solves all equations on all grid levels, as in a patch-based AMR approach.
This means that not only electromagnetic fields and moments are defined on all nodes of all levels, but also macroparticles.
While this may appear as an overhead to deal with macroparticles in coarse regions where there is also a fine mesh and its associated macroparticles, it comes with several advantages.
First, the macroparticles only see one mesh resolution, their shape is thus perfectly constant in time.
Then, merging macroparticles is not required anymore  since macroparticles can simply be deleted as they exit a refined level because their is a self-consistent kinetic flux in the overlapped region of the next coarser level.
As in other AMR PIC codes, refined levels are fed with macroparticles from splitting those from coarser levels at level boundaries.
And once the fine solution is obtained, the electromagnetic field is coarsened onto the next coarser level, and in practice it either overwrites the coarser solution in the overlapped region, or is averaged with it.

To our knowledge, existing AMR PIC codes use in-house developed code for the adaptive meshing mechanism and evolve the system with a time step uniform across all mesh levels \cite{Fujimoto:2008jv, muller2011}.
In such a case, the time step is thus constrained by the finest grid of the model, which leads to much heavier simulations than necessary.
The MLMD method was also originally proposed with uniform time stepping across grid levels. It was then updated to consider a proper stepping per level \cite{innocentiMLMDdt}.
Coarser levels can evolve much fewer cycles than refined ones, which, considering the dispersive nature of kinetic plasma waves, is much more advantageous than codes based on a uniform and fixed time stepping.
Published results\cite{innocentiMLMD1, beckMLMD2, innocentiMLMDjp, innocentiMLMDmomentum,innocentiMLMDdt} however only demonstrate the method with only one refined level, consisting in a single refined patch with a predefined position which is fixed in time.
Such a code is thus useful when the region in which to enhance the resolution is known in advance and does not evolve with time.
In complex systems, where critical small scale regions are moving, appear and disappear, the lack of adaptivity imposes the refinement of a substantial part of the domain, which may become prohibitive.

Contrary to MHD codes, AMR kinetic codes cannot have arbitrarily large mesh spacing. Kinetic plasmas include intrinsic particle scales that need to be correctly resolved even in regions where "nothing" happens.
Solving explicit fully kinetic equations on a mesh much coarser than the electron Debye length is unstable.
Solving hybrid equations with a mesh and time step much coarser than the local ion scales is irrelevant, if not wrong.
Such an upper bound to the coarsest mesh resolution make modeling very large domains expensive even with AMR.
It thus appears interesting to not only consider refining the mesh and the time step, but also the physical formalism that is resolved.
The MLMD method also appears promising in that regard since each refinement level, having its own macroparticles, could in principle be coupled to levels evolving different equations, given that one knows how to transfer information from one to the other.

Coupling a kinetic solver, operating on critical regions, with a fluid solver, evolving less important regions of the domain, has been an important goal over the last decade.
The first coupling between MHD and fully kinetic PIC equations was achieved for local simulations of magnetic reconnection \cite{2007JCoPh.227.1340S, Ishiguro:2010cu, Usami:2013bc} along the inflow direction.
A 2D coupling was later achieved using anisotropic MHD and Hall MHD and implicit fully kinetic equations using the codes BATS-R-US and iPIC 3D \cite{daldorff2014}.
This method was then extended to 3D \cite{tothEPIC2016}. Simulations embedding one or several rectangular full-PIC regions in a global MHD domain were then performed for modeling the Earth's magnetosphere \cite{chen17MHDEPIC}, Ganymede's magnetosphere \cite{zhou19MHDEPIC, Zhou:2020ea}, the Mars' magnetosphere \cite{Ma:2018es}, or Mercury's magnetosphere \cite{ChenYuxi2019MHDEPIC}. The method has later been implemented between iPIC3D and MPI-AMRVAC \cite{2017CoPhC.221...81M, Makwana:2018jd}.

These works provided for the first time a large scale context to otherwise and until then isolated kinetic boxes.
%However they still present the important limitation that the kinetic domains need to be setup in predetermined locations.
However, the choice of using fully kinetic equations, even through an implicit scheme, forces the resolution of some electron particle scales which prevents a kinetic treatment of ions over large scales. 
Coupling fluid equations to hybrid ones would instead allow to cover much broader regions with kinetic ions.

In this paper we present the design and implementation of a new hybrid kinetic PIC code with adaptive mesh refinement inspired from the MLMD method. AMR will make high resolution more affordable for large domains and set the code architecture for future multi-formalisms simulations. 
Section \ref{sec:hybridformalism} presents the hybrid formalism, its physical assumptions and associated equations.
Section \ref{sec:amr} details the AMR method we employ. 
The proposed code named \PHARE{}, handles an arbitrary number of refinement levels and patches which position and shape is adapted dynamically according to the current state of the solution and refinement criteria. 
A refined time stepping is used to accommodate the local Courant–Friedrichs–Lewy  condition \cite{1967IBMJ...11..215C} required by our explicit scheme, and use the largest possible time step on each refinement level.
Section \ref{sec:validation} presents the validation of the hybrid solver with and without AMR.
Section \ref{sec:software} explains the main points of the code implementation, in particular related to the AMR mechanism.
Finally, section \ref{sec:sumpersp} concludes the paper.

\section{Hybrid Particle-in-Cell formalism}\label{sec:hybridformalism}

\subsection{Physical assumptions and model equations}

In this section we present the physical assumptions concerning the plasma dynamics and the relevant mathematical equations of the hybrid formalism. The magnetic field is evolved from the electric field via Faraday's equation:

\begin{equation}\label{eq:faraday}
     \frac{\partial \mathbf{B}}{\partial t}  = -\nabla\times\mathbf{E} 
\end{equation}

The total current density $\mathbf{j}$ is obtained from Ampere's equation where the displacement current is neglected:

\begin{equation}\label{eq:ampere}
\mu_0 \mathbf{j} = \nabla\times\mathbf{B} 
\end{equation}

\noindent where $\mu_0$ is the vacuum permittivity.
The total ion density, $n_i$, is the sum of the density $n_p$ of each ion population $p$, obtained from integration of their distribution function $f_p$:

\begin{equation}
\label{eq:ni}
n_i\left(\mathbf{r},t\right)  =  \sum_p \int f_p\left(\mathbf{r},\mathbf{v},t\right)d\mathbf{v}
\end{equation}

The total ion bulk velocity $\mathbf{u}$ is computed from the particle flux of all populations and the total ion density:

\begin{equation}\label{eq:vi}
\mathbf{u}\left(\mathbf{r},t\right) =\frac{1}{n_i} \sum_p \int \mathbf{v} f_p\left(\mathbf{r},\mathbf{v},t\right)d\mathbf{v}
\end{equation}

The hybrid formalism assumes that the plasma dynamics occurs at spatial and temporal scales for which quasi-neutrality holds, i.e. the total ion charge density $\rho_i  =\sum_p n_pq_p =\sum_p eZ_p$, where $n_p$, $q_p$ are the particle density and the electric charge of the population $p$, respectively, is equal to the electron charge density $en_e$. With this assumption the electron bulk velocity can be calculated from the known electrical current density $\mathbf{j}$, and ion moments:

\begin{equation}
    \label{eq:currentquasineutral}
     \mathbf{v_e} = \mathbf{u} - \frac{1}{en_e}\mathbf{j} 
\end{equation}

The major physical assumption of the hybrid formalism, that is the less justifiable in collisionless systems, is the closure of the electron moment hierarchy at the pressure level. Currently \PHARE{}  uses the isothermal closure: 

\begin{equation}\label{eq:closure}
    P_e = n_e k_B T_e
\end{equation}

\noindent where $k_B$ is the Boltzmann constant.
This closure is the simplest and most commonly used. It assumes that the electron pressure tensor reduces to a simple scalar, $P_e$, proportional to the electron density. The coefficient $T_e$ defines the temperature which is a constant in both space and time. Since the first three electron moments are known, one can use and rewrite the electron fluid momentum equation to calculate an electric field that is consistent with them. 
We further assume time and spatial scales at which the electron fluid is accelerated and thus neglect the bulk inertia in the momentum equation.
The electric field is then given by:

\begin{equation}\label{eq:ohmdissipless}
     \mathbf{E}  = - \mathbf{v}_e\times\mathbf{B} - \frac{1}{en_e}\nabla P_e 
\end{equation}

In practice, eq. \ref{eq:ohmdissipless} can be problematic because it cannot prevent the possibly infinite thinning of current sheets.
In reality, the thickness of any current sheet would be limited by intrinsic electron particle orbit scales, leading to non-negligible non-gyrotropic pressure tensor components and bulk inertia. 
Since we do not include these scales, we need to add terms in eq. \ref{eq:ohmdissipless} to include a dissipative scale controlling the minimal current sheet thickness in our systems. 
Two terms are added, the classical Joule resistive term $\eta \mathbf{j}$, where $\eta$ is constant, which diffuses large scale structures, and the hyper-resistive term $-\nu \nabla^2 \mathbf{j}$, where $\nu$ is constant, which on the contrary is able to dominate over the Hall term and  defines a dissipation-dominated small scale \cite{aunai2013dissip}.
\new{Hyper-resistivity is also, from a numerical standpoint, a efficient way to dissipate possible accumulation of grid scale fluctuations in hybrid-PIC codes.}
The inclusion of these two terms leads to :

\begin{equation}\label{eq:ohm}
     \mathbf{E}  = - \mathbf{v}_e\times\mathbf{B} - \frac{1}{en_e}\nabla P_e + \eta\mathbf{j} - \nu\nabla^2\mathbf{j}
\end{equation}

Finally, in the hybrid formalism, the evolution of the distribution function of each ion population p, used in equations \ref{eq:ni} and \ref{eq:vi}, is described by the Vlasov equation:

\begin{equation}\label{eq:vlasov}
    \frac{\partial f_p}{\partial t} + \mathbf{v}\cdot\frac{\partial f_p}{\partial \mathbf{r}}  + \frac{q_p}{m_p} \left(\mathbf{E} + \mathbf{v}\times\mathbf{B}\right)\cdot\frac{\partial f_p}{\partial \mathbf{v}}  = 0
\end{equation}

\subsection{Macroparticles and their interaction with the mesh}

In practice, equation \ref{eq:vlasov} is not directly solved in PIC codes, but its solution is equivalently obtained by the evolution of a collection of so-called \textit{macroparticles}, which, according to the PIC method, follow the same dynamical equations as real particles:

\begin{eqnarray}
     m_p \frac{d\mathbf{v}}{dt} &=& q_p\left(\mathbf{E} + \mathbf{v}\times\mathbf{B}\right)\label{eq:partv}\\
     \frac{d\mathbf{r}}{dt} &=& \mathbf{v}\label{eq:partr}
\end{eqnarray}

The moments of the ion distribution used in \ref{eq:ohm} are obtained from macroparticles.
The distribution function $f_p$ of an ion population $p$ at a given point in phase $(\mathbf{r}, \mathbf{v})$ space writes:

\begin{equation}
    f_p(\mathbf{r},\mathbf{v}) = \sum_{l=1}^{N_p} S(\mathbf{r}-\mathbf{r}_l)\delta(\mathbf{v}-\mathbf{v}_l)
\end{equation}

where the sum is performed over the $N_p$  macroparticles of the population $p$, $S$ is a symmetric kernel normalized to $1$ over its compact support, and  $\delta$ is the Dirac function.
Macroparticles thus have a finite spatial extent and a unique velocity.
Each of the $N_c$ macroparticles in a cell is assigned a statistical weight $W = V_c n_c/N_c$ representing its contribution to the cell's density $n_c$ of volume $V_c$.
The function $S$ is chosen to be a B-spline $S_\nu$ of order $\nu$.
The particle number density at the mesh point $\mathbf{r}_{ijk}$ is obtained from the $N_c$ macroparticles with position $\mathbf{r}_l$:
%The density of the population $p$ is obtained from $f_p$ in the cell $(i,j,k)$ of volume $V$ is given by

%\begin{equation}
%    n_{ijk}= \sum_{l=1}^{N_p}\frac{1}{V}\int_V %S(\mathbf{r}-\mathbf{r}_l) \mathbf{dr}
%\end{equation}

%the integral over the cell volume is transformed using the top-hat function $\Pi(x)$ into

%\begin{equation} \label{eq:densitypic}
%    n_{ijk} = \sum_{l=1}^{N_p}\int_{-\infty}^\infty  %S(\mathbf{r}-\mathbf{r}_l) %\Pi\left(\frac{\mathbf{r}}{\mathbf{dr}}\right)\mathbf{dr}
%\end{equation}

%The function $S$ is chosen to be a b-spline of order $\nu$. The zeroth order b-spline $S_0(x)$ is the top-hat function $\Pi(x)$, and higher order b-splines are found by the recursive relation 

%\begin{equation}
%    S_{\nu+1}(x) =\int_{-\infty}^\infty S_\nu(x'-x) S_0(x')dx'
%\end{equation}

\begin{equation}\label{eq:densitypic2}
    n_{ijk} =\frac{1}{V_c} \sum_{l=1}^{N_c}W_l S_\nu(\mathbf{r}_{ijk}-\mathbf{r}_l)
\end{equation}

Similarly, the average particle density flux is obtained from particles velocity $\mathbf{v}_l$:

\begin{equation} \label{eq:flux2}
    \mathbf{\Phi}_{ijk} =\sum_{l=1}^{N_p}W_l \mathbf{v}_l S_\nu(\mathbf{r}_{ijk}-\mathbf{r}_l)
\end{equation}

\PHARE{} supports first, second and third order B-splines $S_1$, $S_2$ and $S_3$ : 

\begin{equation}
\label{app:shape_order1}
S_1(x) = \left\{ \begin{array}{ll}
1-|x| & \textrm{ if } |x| \leq 1  \\
    0 & \textrm{ elsewhere }
\end{array} \right.
\end{equation}

\begin{equation}
S_2(x) = \left\{ \begin{array}{ll}
\frac{3}{4} -x^2 & \textrm{ if }  |x| \leq {^1\!/\!_2}\\
\frac{1}{2}\left( \frac{3}{2} + |x|\right)^2 & \textrm{ if }     {^1\!/\!_2} \leq |x| \leq {^3\!/\!_2}\\
    0 & \textrm{ elsewhere }
\end{array} \right.
\end{equation}

\begin{equation}
S_3(x) = \left\{ \begin{array}{ll}
\frac{|x|^3}{2} -x^2 +2/3 & \textrm{ if }  |x|  \leq 1   \\
\frac{4}{3}\left( 1 - \frac{|x|}{2}\right)^3 & \textrm{ if } 1\leq |x| \leq 2 \\
    0 & \textrm{ elsewhere }
\end{array} \right.
\end{equation}

Increasing the B-spline order uses more mesh nodes. It thus  decreases the noise level, at the price of heavier computations and an increased diffusion of gradients that may exist at a scale close to that of the mesh.

\subsection{Normalization of physical quantities}

The code evolves dimensionless quantities that are obtained with the following normalization. The magnetic field and the particle density are normalized by arbitrary constants $B_0$ and $n_0$, respectively. Masses are normalized to the proton mass $m_p$ and charges to the elementary charge $e$. It follows that the time is normalized to the inverse proton gyrofrequency $\Omega_{0} = (eB_0/m_p)$. Velocities are normalized to the proton Alfvén speed in the reference magnetic field and density $V_{A0} = B_0/\sqrt{m_pn_0\mu_0}$. The distances are thus normalized to the ion inertial length $\delta_{i0} = V_{A0}/\Omega_{ci0}$

\subsection{Spatial discretization}
\begin{figure}
    \centering
    \includegraphics[width=0.9\linewidth]{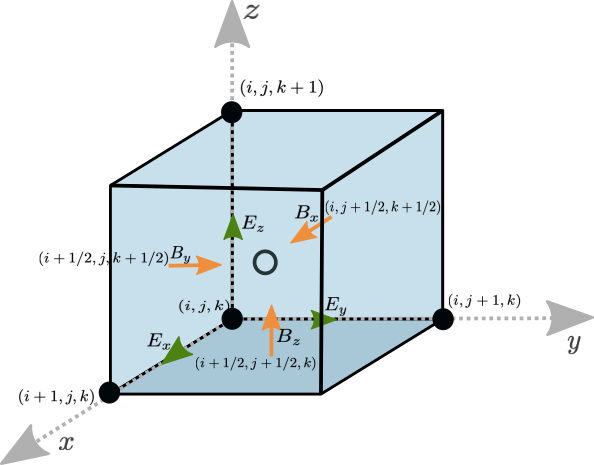}
    \caption{3D representation of the Yee lattice \cite{Yee}. The components of the magnetic and electric fields are represented at their position. The magnetic component are located at primal locations in their direction, and dual locations otherwise, whereas the electric components are on the contrary positioned at dual locations in their direction and primal locations otherwise.}
    \label{fig:yee3D}
\end{figure}

The spatial discretization of equations \ref{eq:faraday} to \ref{eq:ohm} in \PHARE{} follow the Yee layout \cite{Yee}, which preserves the divergence-free character of the magnetic field.
In each dimension, the components of the electric and magnetic field are positioned at integer or half integer multiples of the mesh size, so-called primal and dual positions, respectively.
The 3D version of the Yee lattice is represented on Fig. \ref{fig:yee3D}. By choice, plasma moments are defined on primal positions.

\subsection{Temporal discretization}\label{sec:ppc}
%\andrea{The macro-particle positions and velocities, as well as the electromagnetic fields are advanced in time using the second order....}
We adopt the second order iterated Crank-Nicholson method, with predictor-predictor-corrector stages, as used in the code \texttt{Pegasus} \cite{kunz}.
The process starts at time step $k$ with known electromagnetic fields $\mathbf{B}^k$, $\mathbf{E}^k$, particles position $\mathbf{r}^k$ and velocity $\mathbf{v}^k$, the electron density is obtained from the ion populations density $n_e^k = \sum_p Z_p n_p^k$, and the total ion bulk velocity from the ion population fluxes $\mathbf{u}^k = \sum_p\Phi_p^k/\sum_p n_p^k$.
The first step consists in computing at time $k+1$ the first predicted values (subscript $p1$) of the magnetic field, current density and electric field from Faraday's (eq. \ref{eq:faradayp1}), Ampere's (eq. \ref{eq:amperep1}) and Ohm's (eq. \ref{eq:ohmp1}) laws respectively:

\begin{eqnarray}
   \mathbf{B}^{k+1}_{p1} & = & \mathbf{B}^k - \Delta t\nabla\times\mathbf{E}^k \label{eq:faradayp1}\\
    \mathbf{j}^{k+1}_{p1} & = & \nabla\times\mathbf{B}^{k+1}_{p1} \label{eq:amperep1}\\
     \mathbf{E}_{p1}^{k+1} & = &- \mathbf{u}^k\times\mathbf{B}_{p1}^{k+1}\label{eq:ohmp1}\\  && +\frac{1}{n_e^k}\left(\mathbf{j}^{k+1}_{p1}\times\mathbf{B}_{p1}^{k+1} - \nabla P_e^k\right) \nonumber \\
     & & + \eta\mathbf{j}_{p1}^{k+1}  - \nu\nabla^2\mathbf{j}_{p1}^{k+1} \nonumber
\end{eqnarray}

The time centered value ($k+1/2$) for the predicted electric and magnetic fields is then computed by simple averaging:
\begin{eqnarray}
\mathbf{E}^{k+1/2}_{p1} & = & \left(\mathbf{E}^k + \mathbf{E}^{k+1}_{p1}\right)/2\label{eq:avgp1E}\\
\mathbf{B}^{k+1/2}_{p1} & = & \left(\mathbf{B}^k + \mathbf{B}^{k+1}_{p1}\right)/2\label{eq:avgp1B}
\end{eqnarray}

These time centered fields are then used to "push" the particles, i.e. to compute the particle velocities $\mathbf{v}^k\to \mathbf{v}^{k+1}_{p}$ and positions $\mathbf{r}^k\to \mathbf{r}^{k+1}_{p}$ (the particle pusher algorithm is detailed in section \ref{sec:pusher}). The new values of velocity and positions are then used to calculate the predicted electron density $n_e^{\star k+1}$ and total ion bulk velocity $\mathbf{u}^{\star k+1}$, thus completing the first predictor step.

This is followed by a second prediction, $p2$, of the magnetic field, current density and electric field:

\begin{eqnarray}
 \mathbf{B}^{k+1}_{p2} & =& \mathbf{B}^k - \Delta t\nabla\times\mathbf{E}^{k+1/2}_{p1} \label{eq:faradayp2}\\
\mathbf{j}^{k+1}_{p2} & = & \nabla\times\mathbf{B}^{k+1}_{p2} \label{eq:amperep2}\\
  \mathbf{E}_{p2}^{k+1}  & = & - \mathbf{u}^{\star k+1}\times\mathbf{B}_{p2}^{k+1}  \label{eq:ohmp2}\\ && +\frac{1}{n_e^{\star k+1}}\left( \mathbf{j}^{k+1}_{p2}\times\mathbf{B}_{p2}^{k+1}  -\nabla P_{e}^{k+1}\right)\nonumber \\
  && + \eta\mathbf{j}_{p2}^{k+1} -  \nu\nabla^2\mathbf{j}_{p2}^{k+1}\nonumber
  \end{eqnarray}

The new time centered electromagnetic fields 

 \begin{eqnarray}
  \mathbf{E}^{k+1/2}_{p2} & = & \left(\mathbf{E}^k + \mathbf{E}^{k+1}_{p2}\right)/2\label{eq:avgp2E}\\
\mathbf{B}^{k+1/2}_{p2} & = & \left(\mathbf{B}^k + \mathbf{B}^{k+1}_{p2}\right)/2\label{eq:avgp2B}  
 \end{eqnarray}
  
are then used to push particles velocities  $\mathbf{v}^k\to \mathbf{v}^{k+1}$ and positions $\mathbf{r}^k\to \mathbf{r}^{k+1}$ a second and final time, before computing the moments $n^{k+1}$ and $\mathbf{u}^{k+1}$.

 To finish, a correction step computes the final value of the electromagnetic field and current density:

%(x^{n+1},\mathbf{v}^{n+1})& \longleftarrow & (\mathbf{r}^n,\mathbf{v}^n) \\
%
\begin{eqnarray}
 \mathbf{B}^{k+1} &= &\mathbf{B}^k - \Delta t\nabla\times\mathbf{E}^{k+1/2}_{p2} \label{eq:faradaycor}\\
     \mathbf{j}^{k+1} & = & \nabla\times\mathbf{B}^{k+1}\label{eq:amperecor}\\
      \mathbf{E}^{k+1}  &= &- \mathbf{u}^{k+1}\times\mathbf{B}^{k+1} \label{eq:ohmcor}\\ && +\frac{1}{n_e^{k+1}}\left(\mathbf{j}^{k+1}\times\mathbf{B}^{k+1}- \nabla P_e^{k+1}\right)  \nonumber \\
      & & + \eta\mathbf{j}^{k+1} - \nu\nabla^2\mathbf{j}^{k+1}\nonumber
\end{eqnarray}

\subsection{Particle pusher}\label{sec:pusher}

Macroparticles are pushed following the Boris algorithm \cite{Boris1970}, here  split into the 5 following steps. The particle positions and velocities are defined at the same time $k$ but the predictor-predictor-corrector scheme uses the electromagnetic fields defined at the time $k+1/2$, so macroparticles are first push half a time step:

\begin{equation}\label{prepush_eq}
 \mathbf{x}^{k+1/2} =  \mathbf{x}^{(k)} + \frac{\Delta t}{2} \mathbf{v}^{k} 
\end{equation}

The electric $\mathbf{E}^{k+1/2}$ and magnetic field $\mathbf{B}^{k+1/2}$, known on the mesh, are interpolated at the particles' position using the B-spline assignment functions discussed previously. These interpolated fields $\mathbf{\tilde{E}}^{k+1/2}$ and $\mathbf{\tilde{B}}^{k+1/2}$ are then used to update the particles' velocities following:

\begin{equation}\label{boris_eq1}
\mathbf{v}^-  =  \mathbf{v}^{k} + \frac{\Delta t }{2m} \mathbf{\tilde{E}}^{k+1/2}
\end{equation}

\begin{equation}\label{boris_eq2}
    \mathbf{v}^+  =  \mathbf{v}^- + \frac{q\Delta t}{m} \left( \frac{\mathbf{v}^- + \mathbf{v}^+}{2}\right)
\times \mathbf{\tilde{B}}^{k+1/2}
\end{equation}

\begin{equation}\label{boris_eq3}
    \mathbf{v}^{k+1}  =  \mathbf{v}^+ + \frac{\Delta t}{2m} \mathbf{\tilde{E}}^{k+1/2}
\end{equation}

The final particles' position is then given by:

\begin{equation} \label{postpush_eq}
    \mathbf{x}^{k+1}  =  \mathbf{x}^{k+1/2} + \frac{\Delta t}{2} \mathbf{v}^{k+1}
\end{equation}

\section{Adaptive Mesh Refinement}\label{sec:amr}

\subsection{Patch hierarchy}
The adaptive mesh refinement in \PHARE{} follows the so-called \textit{patch-based} approach.
Equations are solved in rectangular domains of uniform mesh resolution called \textit{patches}.
Each patch contains data that is either represented on a grid (e.g field quantities) or that is gridless (e.g macroparticles).
Because of spatial derivatives and of the finite extent of the macroparticle assignment function $S_\nu$, computing the solution in the patch domain requires the knowledge of the solution outside the patch in a so-called \textit{ghost layer}.
This ghost layer represents the current state of the solution of the adjacent regions of the simulation domain, and its thickness depends on whether it concerns fields or macroparticles, as explained in more details in sections  \ref{ss:ghost_particles} and  \ref{ss:field_ghosts}.

\begin{figure}
    \centering
    \includegraphics[width=0.9\linewidth]{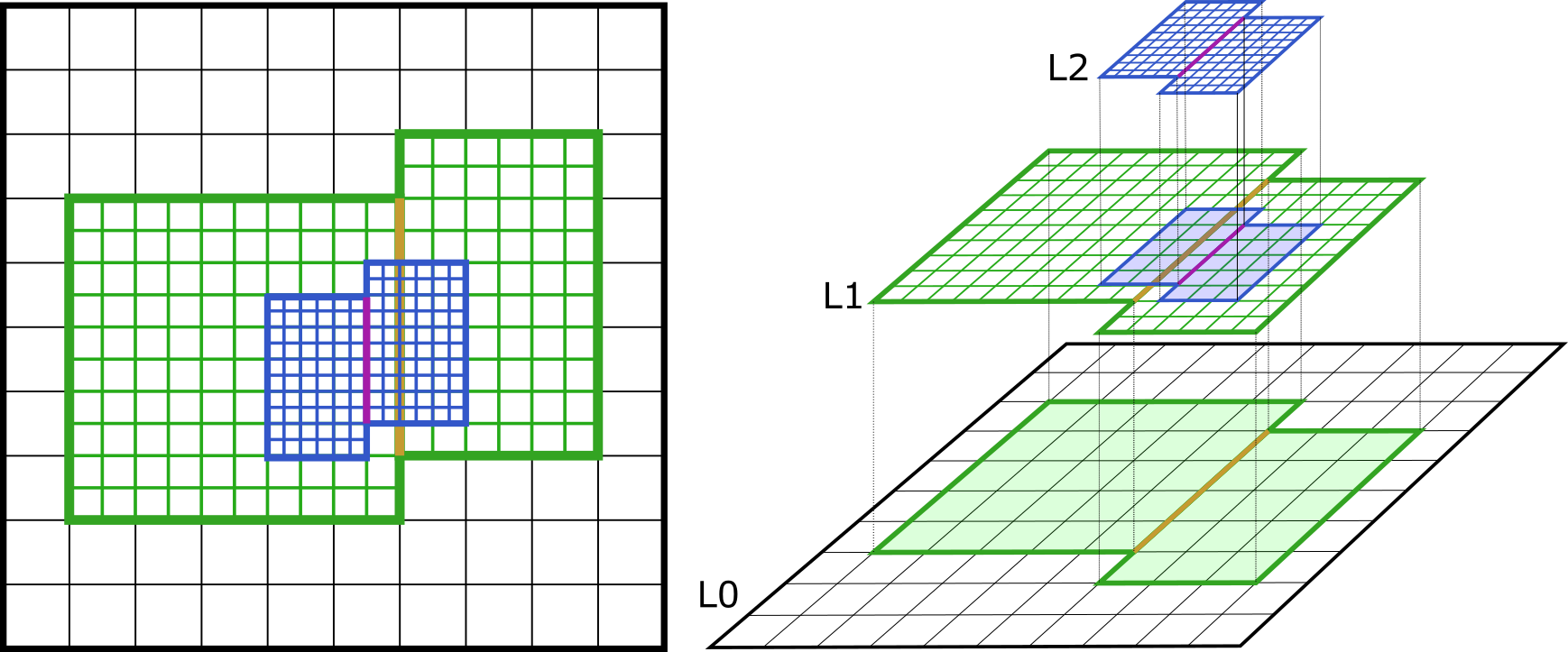}
    \caption{Illustration of a patch hierarchy composed of 3 levels. In this simplified example, the coarsest  level $L_0$ is composed of a single patch represented in black, $L_1$ has two patches represented in green. Finally $L_2$ has two blue patches. The two patches on $L_1$ (resp. $L_2$) are separated by a patch boundary represented in orange (resp. purple). The other boundaries of $L_1$ (resp. $L_2$) are represented in green (resp. blue) and define the so-called level boundaries.}
    \label{fig:patchhierarchy}
\end{figure}

The union of all patches of the same mesh resolution defines a so-called \textit{patch level}, and the ensemble of patch levels forms the so-called \textit{patch hierarchy}. This is represented on figure \ref{fig:patchhierarchy}, in a two-dimensional geometry, for illustrative purposes.
Each level increment corresponds to a spatial refinement of the mesh of the enclosed patches by a specific refinement ratio. \PHARE{} adopts an isotropic refinement ratio set to 2.
The number of patch levels composing the patch hierarchy can change over time depending on the state of the system, and is bounded by a runtime user supplied parameter.

Cells constituting the ghost layer of a patch can be of two types, represented on Fig. \ref{fig:patchhierarchy}.
Cells overlapping those of the domain of a neighbor patch in the same patch level are called \textit{patch ghost cells}.
The information (field and particles) in these patch ghost cells is just an exact copy of the information existing in the overlapped neighbor domain cells.
Cells overlapping the next coarser level mesh are called \textit{level ghost cells}. Their data is obtained from refining next coarser level data.
\emph{Patch ghost cells} overlap the cells of the domain of a neighbor patch in the same patch level. The information (field and particles) in these patch ghost cells are just an exact copy of the information existing in the overlapped neighbor domain cells. \emph{Level ghost cells} overlap the next coarser level mesh and their data is obtained from refining next coarser level data.

By definition, the union of the patches of the coarsest patch level cover the whole simulation domain. Therefore ghost layer cells of that level are always patch ghost cells (for periodic domains).
The filling of field quantities in ghost nodes and macroparticles in ghost cells for those two types of boundaries need to be handled differently, the details are given in subsections \ref{ss:ghost_particles} and \ref{ss:field_ghosts}.

\subsection{Refined time stepping evolution}
\new{In a hybrid code (or any code that allows the inertial decoupling of the ion flow from the magnetic field), the most severe constraint for the time step originates from the dispersive propagation of whistler waves, for which the frequency $\omega$ follows $\omega \sim k^2$.
In such a code, the time step $\Delta t$ must accommodate the propagation of fastest whistler waves, i.e. those at the smallest wavelength possible on a given mesh.
In the context of an AMR code, where the mesh size is incrementally refined, two solutions are possible.
The first, consists in having a unique time step throughout the mesh hierarchy.
This unique time step must thus satisfy the constraint brought by the finest mesh at a given time.
Considering the whistler dispersion imposes $\Delta t \sim \Delta x^2$, this means, for a hierarchy composed of 3 levels, that the time step needs to be 16 times smaller on the coarsest level (covering the entire domain) that what the whistler dispersion would impose.
A much better solution, that we adopted here, consists in having a time step per mesh level.
In this case, the time step must be small enough for propagating the fastest whistler waves on the coarsest level and incrementally divided by $r^2$ as the mesh size is divided by $r$.
}
\PHARE{} advances the solution on the patch hierarchy using \new{such} a refined time stepping procedure, thus avoiding the evolution of the whole domain being constrained by the smallest time step associated with the finest level.

\begin{figure}
    \centering
    \includegraphics[width=0.7\linewidth]{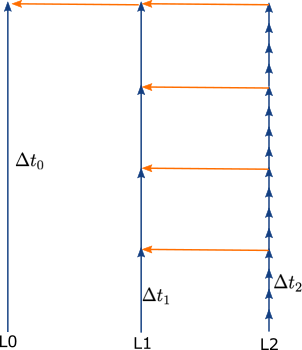}
    \caption{Refined time stepping evolution of the patch hierarchy illustrated for 3 patch levels. On that figure, the evolution of the coarsest ($L_0$), intermediary ($L_i$) and finest level ($L_3$) is represented from left to right. Blue vertical arrows represent the time step taken by each level. Horizontal orange arrows represent the coarsening of $L_{i+1}$ data onto $L_{i}$ mesh.}
    \label{fig:recursivestepping}
\end{figure}

 The time stepping is illustrated in Fig. \ref{fig:recursivestepping} for a patch hierarchy composed of 3 patch levels.
 The evolution proceeds in a recursive manner as follows.
 Level $L_0$ advances the solution on all of its patches by one coarsest time step $\Delta t_0$.
 Then before it advances another step, $L_1$ advances its patches by $\Delta t_1 = \Delta t_0/r^2$ were $r$ is the refinement ratio.
 Then since $L_2$ is the finest level at that time, it is advanced $r^2$ steps, of size $\Delta t_1/r^2$, up to the time reached by $L_1$.
 At this point, the solution obtained on $L_2$ has the best resolution and thus is coarsened to overwrite the one existing on $L_1$ on overlaped regions, before $L_1$ is advanced by $\Delta t_1$ again.
 This is repeated $r^2$ times until $L_1$ itself reaches the time at which $L_0$ is and coarsens its solution onto $L_0$, which then advances to its second step, etc.
 This procedure is repeated as many times necessary until $L_2$ reaches the final desired simulation time.

\subsection{Initialization of a patch level}
This section explains how data is initialized on a patch level, depending on whether it concerns the initialization of the coarsest level of the hierarchy, a new finest refined level, or the reorganization of an existing one.

\subsubsection{Initialization of the coarsest level}
 The coarsest level uses user-defined functions to initialize fields and particles at the start of a simulation.
 The macroparticles are loaded on a per-cell basis according to a user supplied density profile and a constant number of macroparticles per cell and per population.
 The velocity of macroparticles is distributed according to a Maxwellian distribution function with moments locally defined via the user supplied functions.
 Once the magnetic field and the plasma moments are set, the electric field simply results from Ohm's law (eq. \ref{eq:ohm}).

\subsubsection{Initialization of a new refined level}\label{sec:initreflev}
Whether it is at the start time of the simulation or when a new level at the finest resolution is being created during the simulation, a refined level is initialized by refining data existing on the next coarser level.
\new{Macroparticles are refined from the aforementioned splitting procedure proposed in Smets et al.\cite{Smets:2021dw}
The method uses pre-computed and tabulated values for the relative position and weights of the refined macroparticles.
Plasma moments are not refined since they can be rather more consistently obtained from the newly obtained macroparticles.
The electric field needs to be refined as it will be required for solving Faraday's equation \ref{eq:faradayp1}.
It is obtained by linear interpolation from the coarse mesh by taking the coarse value if the fine node falls on top of the coarse one, or the average of the surrounding coarse values otherwise.
In 1D for the sake of simplicity, a fine primal node thus either takes the value of the overlapped coarse node if any, or the half sum of the two surrounding (and equidistant) nodes.
A dual fine node will take a linear combination of the two surrounding coarse dual values, weighted either $(1/4, 3/4)$ or $(3/4,1/4)$ depending on whether the fine dual node is located on the first or second half of the coarse cell, respectively.
The magnetic field is obtained by propagating the coarse values in a way that preserves its divergence free property.
To do this, we impose the magnetic flux across a coarse face to be conserved on the fine enclosed faces by setting the component across each of the enclosed fine faces equal to that of the coarse face.
}

%Fields are refined using linear spatial interpolation while macroparticles are refined using the %splitting procedure described in Smets et al.\cite{Smets:2021dw}.
%This splitting procedure ensures that the combination of the refined macroparticle b-splines is either exactly or optimally close to the b-spline of the original coarse macroparticle, and thus preserves the contribution of macroparticles to the mesh.
%Unlike other methods (see \cite{Smets:2021dw} and references therein), it ensures that no or minimal spurious noise is added to the refined grid.
%\rsm{Unlike previous methods (see \cite{Smets:2021dw} and references therein), this splitting procedure ensure that the assignment function of a split set of (fine) particles is exactly equal or optimally conserves the assignment function of the generating set of (coarse) particles without any grid consideration.}

\subsubsection{Initialization of a level during regridding}
The last way a level can be initialized is when regridding occurs. Regridding is the change of the patch hierarchy structure in terms of patch number, position and  geometry, and it is often employed to improve the distribution of spatial resolution around critically important regions as they evolve.

\begin{figure}
    \centering
    \includegraphics[width=0.8\linewidth]{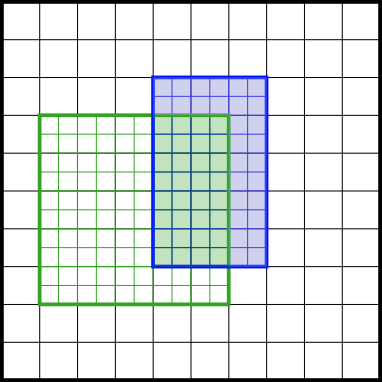}
    \caption{Illustration of a regridding operation in a patch hierarchy with two levels. The green level is being removed and replace by the blue one. The new level copies data from the red cells of the old level and refines data from coarser level in the blue cells.}
    \label{fig:regridding}
\end{figure}

Whenever regridding occurs, a set of levels are removed from the patch hierarchy, and replaced by new ones, as illustrated in Fig. \ref{fig:regridding}. Data is copied (resp. streamed) from local (resp. remote) memory where there is an overlap between old and new patch level regions, and refined from coarser data where there is none. In the latter case, the refinement of fields and particles is performed through the same functions as those used for the initialization of new finest level initialization.

\subsection{Macroparticle deposition to the grid and the handling of ghost particles at patch and level boundaries}\label{ss:ghost_particles}
%A macroparticle represented by a b-spline of order $p$, interacts with $p+1$ grid nodes. Thus, the layer containing ghost macroparticles is 1 (resp. 2) cell(s) wide for $p=1$ (resp. $p=2,3$).
%For fields, there are $(p + l)/2$ ghost nodes, where $l=1$ (resp. $l=0$) if the field is located on dual (resp. primal) nodes in the direction considered.\todo{revise these numbers...}

We now describe how the moments of the ion distribution function are computed on the grid from the macroparticles and introduce the types and use of ghost macroparticles.
Without loss of generality an example for a 1D grid and second order interpolation is illustrated in Fig. \ref{fig:momentghosts}.
The first step in the computation of the moments consists in depositing the properties of each macroparticle located in patch domain cells.
This is the step \textbf{A} of Fig. \ref{fig:momentghosts}. 
At this point, nodes 2 to 5 are so-called \textit{complete}, in a sense that all macroparticles that can reach them have contributed to them. 
For the case of second order interpolation, nodes 0 and 1, as well as nodes 6 and 7, however, are \textit{incomplete}.
They received contributions from domain macroparticles,  but lack that of macroparticles located outside the domain. 
This missing contributions is the main reason why there is a need for so-called \textit{ghost} macroparticles, i.e. macroparticles located beyond the patch domain boundaries that represent the ion populations in these regions. 
Another reason is that one needs to inject macroparticles into the patch domain consistently with the particle flux existing at the boundaries. This said, we shall now see that two types of ghost macroparticles exist in \PHARE{}: patch  and level ghost macroparticles, as described in the two following paragraphs.

\subsubsection{Patch ghost macroparticles}

Patch ghost macroparticles are those located in patch ghost cells, i.e. those overlapping domain cells of a neighbor patch in the same patch level.
By construction, these macroparticles are clones of those positioned in neighbor patches domain cells overlapping the ghost layer.
Like those in the domain cells, patch ghost macroparticles are pushed.
Those entering the domain contribute to the moments on the mesh like domain macroparticles in step \textbf{A} and are copied into the domain particle array.
 Because domain particles of all neighbor patches have moved, at this point the ensemble of macroparticles in the patch ghost layer is not the exact copy of the neighbor overlapped domain cells anymore.
 Patch ghost layers are thus refilled from neighbor patches, through either a copy from local memory if the patch and the neighbor are on the same MPI process, or a transfer if they are not.
 Note that for interpolation of order 2, represented on Fig. \ref{fig:momentghosts}, ghost macroparticles located as far as $1.5\Delta x$ away from the border can contribute to domain nodes. For interpolation of order 1 and 3, the ghost layer width becomes $\Delta x$ and $2\Delta x$, respectively.
 Once received, patch ghost particles are interpolated onto the grid, which in Fig. \ref{fig:momentghosts}, completes nodes 0 and 1 and is illustrated as step \textbf{B} in the figure.
 
 \subsubsection{Level ghost macroparticles}\label{ref:lvlghostpart}
 %After domain macroparticles and patch ghost particles are pushed, only those which remained or came into the domain see their contribution to the moments deposited on the grid. The handling of macroparticles leaving the patch domain area after having been pushed depends on which step of the predictor-predictor-corrector is concerned. The first push updates the positions and velocities in a temporary buffer to preserve the original state for the second push, and discards leaving particles for the deposit phase. However the second push changes the particle position and velocity buffer in place and deletes leaving ones. Patch ghost macroparticles entering the domain are copied into the domain macroparticle buffer only at the second push.
 
 %Once this is done, ghost nodes overlapping neighbor patch domain nodes of the same level, and close-by domain nodes, lack contributions from macroparticles that would be outside the domain on neighbor patches. Copy (resp. streaming) of neighbor patches macroparticles from local (resp. remote) memory is then performed. This copy of patch ghost particles from neighbor patches is performed once domain particles have been pushed so that the ghost layer macroparticles are indeed the clones of neighbor domain ones. Once received, patch ghost macroparticles deposit their contribution onto the grid, which completes the deposition phase on patch ghost nodes and close-by domain nodes.
 
\begin{figure}
    \centering
    \includegraphics[width=\linewidth]{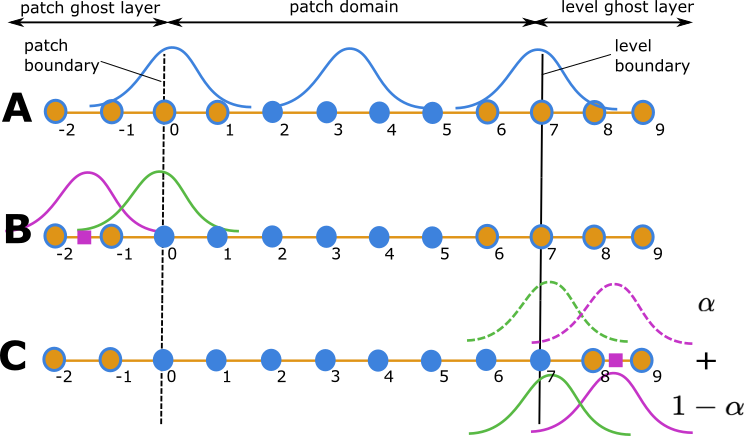}
    \caption{Representation of three steps \textbf{A}, \textbf{B} and \textbf{C} of the moment deposition phase from macroparticles to the mesh of a patch. For the example, this patch is one that is, on its left, adjacent to another patch of the same level, and at the border of the level on its right. Left and right boundaries are thus a patch boundary and a level boundary, respectively. On each step, primal nodes are represented by colored circles. Nodes colored in blue have, at the given step, received the contribution of all macroparticles that can reach them, and are said "complete". Orange nodes circled in blue have received some but not all contributions and are thus said "incomplete". 
    Macroparticles are represented by their shape factor, here at order 2. on step \textbf{A}, blue particles represent domain particles. On step \textbf{B}, the green and purple macroparticles represent patch ghost macroparticles that are the closest and farthest from the domain, respectively. The purple particle is just closer to the limit of $1.5dx$ visualized by the purple square, beyond which no particle at this interpolation order can reach a domain node. On step \textbf{C}, the color code green and purple as the same meaning. There are however 4 macroparticles, representing the refined macroparticles obtained from splitting next coarser ones at its previous time (solid lines) and current time (dashed lines). $\alpha$ and $1-\alpha$ are linear time interpolation coefficients weighting the contribution of the future and past sets of macroparticles, respectively.}
    \label{fig:momentghosts}
\end{figure}

At the end of step \textbf{B} of Fig. \ref{fig:momentghosts}, nodes 6 and 7 thus still lack contributions.
This other boundary of the patch is a level boundary.
There are therefore no neighbor patch cells from which to clone macroparticles in that ghost layer. 
The kinetic state in this region is only represented in the next coarser level and that is where the information is thus taken from. 
The level ghost cells are therefore filled by splitting next-coarser macroparticles, to ensure the continuity of phase space at these boundaries.

It is worth noting that at the time $t$ these operations are performed on level $L_{i}$, the next coarser level $L_{i-1}$ already is at a later time, and there is no representation of its distribution at time $t$.
\PHARE{} handles this situation with two different level ghost macroparticle buffers.
One represents macroparticles refined from next coarser ones at its previous step, the other represents those refined from the current next coarser state, i.e. future time for level $L_i$.
Nodes 6 and 7 receive the contribution of both buffers, each being weighted via linear time interpolation between the two surrounding next coarser times. 
This completes nodes 6 and 7 and represents step \textbf{C} of Fig. \ref{fig:momentghosts}.

Like for patch ghost boundaries, the domain must also receive incoming macroparticles at level boundaries.
The two level ghost macroparticle buffers decribed above represent fixed times and cannot change during all steps taken by level $L_i$ to reach the time at which level $L_{i-1}$ is.
A third buffer is thus used.
At the start of the substepping cycle of level $L_i$, i.e. when level $L_i$ is at the same time as level $L_{i-1}$, this buffer is identical to the other buffer defined at this time.
However, macroparticles contained in that buffer are pushed at each level $L_i$ substep.
Those macroparticles which enter the patch domain are deposited on domain nodes, exactly like is done for entering patch ghost macroparticles.

Since each level has its own set of macroparticles, those which leave a patch domain after being pushed, are simply deleted.
In the case they are leaving through a patch boundary, this means that its clone has entered the neighbor patch domain from the patch ghost layer.
When a macroparticle leaves through a level boundary, nothing is done except deleting the macroparticle, this kinetic flux being already represented consistently on the next coarser level.

\subsubsection{Number of ghost cells for particles}
The number of ghost cells filled with macroparticles depends on the interpolation order.
At interpolation order 1, macroparticles interact with its two surrounding primal or dual nodes.
The border primal node thus cannot be reached by macroparticles outside the domain further than one cell away.
We thus only need one ghost cell filled with macroparticles.
At order 2, macroparticles as far as 1.5 cell away from the domain can reach the border primal node.
The particle ghost layer is thus chosen to be two cells wide.
At order 3, two cells need to be filled with macroparticles and thus the width of the ghost layer is the same as for interpolation order 2.

\subsection{Filling field ghost nodes}\label{ss:field_ghosts}

%After the update of the magnetic field, electric field, and current density, the ghost nodes need to be filled with appropriate values.
\new{After their update in the domain, quantities in the code need to be known on ghost nodes.}
On patch ghost nodes, data is just copied (resp. streamed) from local (resp. remote) memory from overlapped nodes on the same patch level.
However for level ghost nodes, values need to be obtained from the next coarser level.
\new{Care must be taken when carrying data from the next coarser level since, because of the refined time stepping procedure, coarse data is not always defined at the time needed on the fine ghost node, but already reached a future time.
After Ampere's law is computed, the current density $\mathbf{J}$ is needed on ghosts nodes because of the Laplacian operator in the hyper-resistive term in Ohm's law eq. \ref{eq:ohm}.
To fill on level ghost nodes, we first perform a linear combination of the previous and future states of the next coarser level. The interpolation coefficient simply is the ratio between the time elapsed since the last level synchronization and the next coarser time step.
Since the fine level takes 4 sub-steps within a coarse time step, the coefficient takes values in $(1/4, 1/2, 3/4, 1)$.
This time interpolated state is then refined spatially as explained in section \ref{sec:initreflev}.}

\new{Macroparticles exist in level ghost cells over a distance calculated for them to complete the moments on domain nodes, as explained in section \ref{ref:lvlghostpart}.
If level ghost nodes receives contributions from these particles too, they miss contributions from the plasma that exist beyond this ghost particle layer, and are therefore not completed.
These ghosts nodes, needed for the genericity of the coarsening stencil, are thus obtained with the same space and time interpolation procedure as the current density.}

\new{Due to the structure of the Yee mesh, knowing the electric field components everywhere on cell edges is enough to update the magnetic field on cell faces.
Therefore, we set the values for the electric field on level ghost nodes, but not for the magnetic field.
The filling of electric field level ghost values is, however, not performed with the aforementioned space/time interpolation.
Special attention is needed to ensure that the self-consistent evaluation of Faraday's law on level border faces ends up, at the end of the substeps taken by the level, to the exact same flux reached by the next coarser level on the shared coarse cell face, and thereby ensure flux conservation across cell faces shared between coarse and fine levels during that time.
This is done following the idea of Loring et al. \cite{2008ASPC..385..158L}, that consists in decomposing the induction that occurred over the coarse step, into equal sub-inductions on the shared fine faces.
During each substep taken by a fine level, the circulation of the electric field around the union of border fine cells enclosed in a coarse cell, is thus set equal to the circulation along the same contour during the coarse step. Multiplied by a quarter of the coarse time step at each fine step, this leads to the same magnetic flux evolution at the end of the four substeps.
In our case, the magnetic flux on the next coarser level has last been updated with $\mathbf{E}_{p2}^{k+1/2}$ in eq. \ref{eq:avgp2E}.
This value is thus saved and interpolated spatially (and not temporally) on the fine level ghost nodes following the method employed in Fujimoto et al. \cite{fujimoto2011}, after each evaluation of the Ohm's law during the fine sub-stepping.
}

The total number of field ghost nodes is constrained by the width of the macroparticle ghost layer and the fact that these ghost particles are pushed and thus need to interact with mesh nodes.
At first order interpolation, we thus need two primal and dual ghost nodes.
The same reason implies having 3 and 4 ghost cells for fields of orders 2 and 3, respectively.

\subsection{Coarsening of the fine solution}

\begin{figure}
    \centering
    \includegraphics[width=1.0\linewidth]{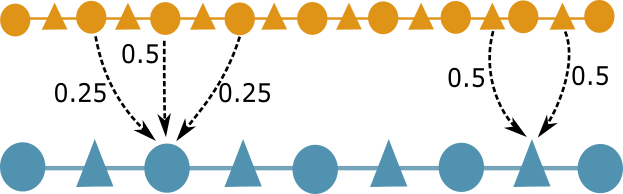}
    \caption{Illustration of the coarsening algorithm on a coarse (blue) patch from refined (orange) patch values. Circles and triangles represent primal and dual positions, respectively. }
    \label{fig:coarsening}
\end{figure}

The assumption behind using AMR is that the solution gets better as the mesh gets finer.
Thus once a fine patch level has reached the same time as its next coarser level, the coarse region overlapped by the fine level is overwritten by the fine solution via a coarsening operation.
Note that in \PHARE{} this coarsening only concerns meshed quantities and not particles.
The particle distribution function in a patch level is \textit{not} coarsened onto its next coarser level. 
Macroparticles located in regions overlapped by a finer level always see electromagnetic fields updated by the finer solution before taking a new step.
This ensures the dynamics is consistent throughout levels and the whole hierarchy stays synchronized.
\new{The coarsening intervenes whenever a fine level reaches the same time as the next coarser one, as illustrated by orange arrows on Fig. \ref{fig:recursivestepping}.
This means that each level except the finest one, will always see an update of its solution in some region before taking another step.}
\new{The magnetic field is coarsened in a way constrained by its divergence-free property.
Specifically, the magnetic flux through a coarse face is set equal to the total fine flux through the enclosed fine faces.
This is ensured by setting each component of the magnetic field on its coarse face equal to the average of those defined on the fine faces.}
The coarsening of other fields depends on whether the quantity is positioned at a dual or primal location.
Dual coarse nodes receive the equally weighted contribution of the two surrounding fine dual values while primal nodes receive the weighted (0.25, 0.5, 0.25) contribution of the three surrounding fine primal values.
This operation is represented along a single dimension in figure \ref{fig:coarsening}, 2D and 3D coarsening just consist in the outer product of this pattern.
\new{We have found the coarsening of the electric field  to produce no noticeable effect on the solution.
The reason is that the electric field does not follow an evolution equation, but is instead assigned by the Ohm's law and is not a true variable of the system but merely an intermediate one.
More specifically, the effect of coarsening or not the electric field is limited to the first evaluation of Faraday's equation in the first predictor step (eq. \ref{eq:faradayp1}), since the electric field is anyway completely re-assigned to a new value when next evaluating the Ohm's law in eq. \ref{eq:ohmp1}.}
\new{The coarsening of the moments, however, is crucial, precisely because the electric field is completely re-assigned by eq. \ref{eq:ohmp1}.
Not coarsening moments would lead the electric field on the coarse mesh to feel the finer solution only through the coarsened magnetic field and ensure a weak coupling between levels.
Updating the coarse moments, on the contrary, allows the coarse electric field to be consistent with the dynamics found on the finer levels and subsequently impose a consistent force on the coarse population, thereby ensuring synchronization between the dynamics on different levels.}

%This strategy was adopted to use the most information possible from the fine grid although strictly speaking it does not prevent the aliasing of the fine level Nyquist frequency, which possible effect deserves an investigation beyond the scope of the current paper.
%In the results presented in the current study, although implemented as well for the ion density and bulk velocity fields, only the fine electromagnetic fields are coarsened onto the next coarser level. 

\subsection{Refinement strategy}
Periodically, the patch hierarchy needs to be updated for the fine levels to map critical regions of the solution.
A criterion is therefore defined to "tag" domain cells that should be part of a region to be refined.
In general, refinement is needed in regions where the grid is insufficiently fine to resolve the fine structures that would self-consistently appear in the solution.
The need for refinement is usually estimated in an empirical manner.
For instance the AMR MHD code \texttt{Flash} \cite{fryxell2008} uses by default an adaptation of the Lohner \cite{lohner87} criterion, an estimator based on a modified second derivative, normalized by the average of the gradient over one computational cell.
Such criterion has been tested in our code but leads to many regions being sporadically refined as a result of the great sensitivity of this method to the numerical noise in the PIC solution.
There is in fact no unique recipe to design a tagging criterion.
One could even argue that the best set of criteria depends on the specific application of the code the user has in mind.
The code \texttt{Flash} provides a way for user to modify or define a custom refinement criterion.
The AMR electromagnetic full PIC code in \cite{fujimoto2006} tags cells depending on their size relatively to the local Debye length because it is a severe spatial constraint on the stability of explicit full PIC codes and also on the local amplitude of the electron current, a criterion fine-tuned for the study of magnetic reconnection.
In the AMR hybrid PIC code \texttt{AIKEF} \cite{muller2011}, the local current density is said to "guide" the refinement but the exact criterion is not specified.
In this paper our goal is to provide an example of a criterion that allows to track sharp gradients and that serves to validate our AMR implementation, it is not our aim here to provide a universal tagging recipe.
In the future, \PHARE{} will also provide a way to define custom refinement criteria.

\new{In this study, we use a criterion defined empirically that quantifies the local variation of the magnetic field at cell $ijk$ } :

\begin{equation}\label{eq:tagging}
%\sigma_\alpha^q = \frac{u_{i+2} - u_i}{1 + u_{i+1} - u_i}
r_{ijk} =\max_{\substack{d=\left\{i,j,k\right\}\\\alpha = \left\{x,y,z\right\}}} \frac{|B_\alpha^{d+2} - B_\alpha^{d}|}{1 + B_\alpha^{d+1} - B_\alpha^d }
\end{equation}

\noindent
the max is taken along each dimension independently, and then for each magnetic field component. 
 
Finally, the cell tag is set to $1$ or $0$ depending on whether the ratio $r$ exceeds an empirical threshold $\sigma$ or not. Results in this paper use $\sigma=0.1$. 

\begin{equation}\label{eq:tagging_threshold}
c = \left\{
    \begin{array}{ll}
        1 & \mbox{if } r_{ijk} > \sigma \\
        0 & \mbox{otherwise}
    \end{array}
\right.
\end{equation}

Once cells are tagged for refinement, they are grouped into boxes by a so-called tile clustering algorithm \cite{gunney16}, used to adapt the number and geometry of patches in a patch level.

Section \ref{sec:movingtd} will validate this approach is able to capture gradients and show its ability to capture fine structures.

%\todo{consider also explaining that although tagging criteria is cell-based, this is patch-based\caption{Table to test captions and labels.}

\section{Validation of the code}\label{sec:validation}

A large number of unit and functional tests have been developed and are executed at each modification of the code. These tests are focused on the behavior of specific components of the code, from an engineering point of view. They decrease a lot the probability components are malfunctioning and above all ensure successive modifications do not alter the working state of the code.

In addition to these tests, we also perform a series of validation tests on physical processes, which are ran during nightly builds when the code has changed.
%We now detail these validation tests. We first discuss tests validating the hybrid solver without refinement, and then with refinement, static and dynamic.
We now discuss these validation tests by starting first with those validating the hybrid solver without refinement, and then including static and dynamic refinement.

\subsection{Validation of the hybrid solver}
\subsubsection{Wave dispersion relations}

\begin{figure}
    \centering
    \includegraphics[width=\linewidth]{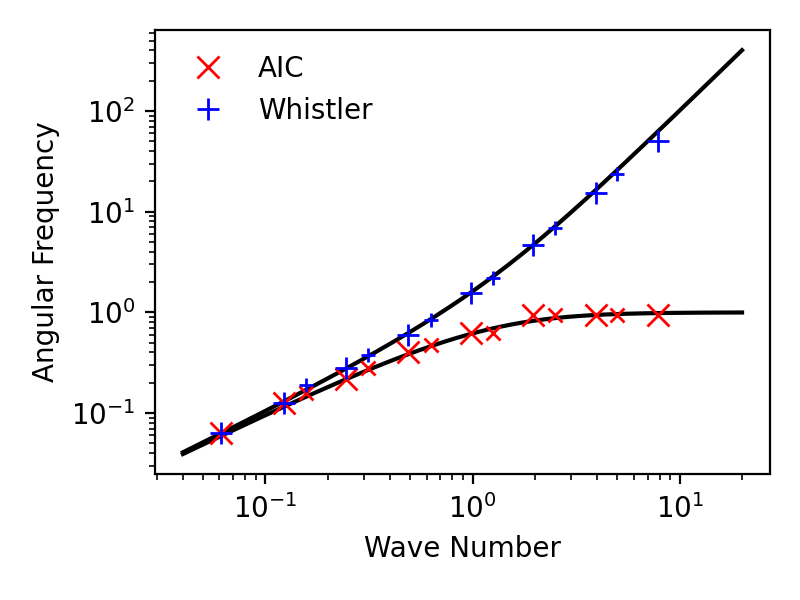}
    \caption{($\omega, k$) diagram for 1D (larger crosses) and 2D (smaller crosses) computations of the Alfvén-Ion-Cyclotron (red) and Whistler (blue) modes displayed in log-log scales.}
    \label{fig:wk}
\end{figure}

%\begin{table}
%\begin{center}
%\begin{tabular}{ |p{2cm}||p{2cm}|p{2cm}|p{2cm}|p{2cm}|p{2cm}|  }
% \hline
% &\multicolumn{2}{|c|}{1D} & & \multicolumn{2}{|c|}{2D} \\
% \hline
%  & low & high & & low & high\\
% \hline
% $n_x$ & 128 & 32 & ($n_x,n_y$) & (200,10) & (100,10)\\
% \hline
% $d_x$ & 0.8 & 0.2 & ($d_x,d_y$) & (0.4,0.4) & (0.2,0.2)\\
%  \hline
% $\Delta t$ & 0.01 & 0.001 &  & 0.00125 & 0.00025\\
% \hline
% T & 200 & 20 &  & 200 & 20\\
% \hline
% $m$ &\multicolumn{2}{|c|}{1,2,4,8} &  & {2,4,8} & {4, 8, 16} \\
% \hline
%\end{tabular}
%\end{center}
%\caption{Table to test captions and labels.}
%\label{table:dispersion}
%\end{table}

\begin{table}
\begin{center}
\begin{tabular}{ p{2cm} p{2cm} p{2cm} }
 \hline
 \hline
 \multicolumn{3}{c}{dim=1} \\
 \hline
 setup & "low" & "high" \\
 $n_x$ & 128 & 32 \\
 $d_x$ & 0.8 & 0.2 \\
 $\Delta t$ & 0.01 & 0.001 \\
 T & 200 & 20\\
 $m$ & (1,2,4,8) & (1,2,4,8) \\
 \hline
 \multicolumn{3}{c}{dim=2} \\
 \hline
setup & "low" & "high" \\
 $(n_x, ny)$ & (200, 10) & (100, 10) \\
 $(d_x, d_y)$ & (0.4, 0.4) & (0.2, 0.2) \\
 $\Delta t$ & 0.00125 & 0.00025 \\
 T & 200 & 20\\
 $m$ & (2,4,8) & (4,8,16) \\
 \hline
 \hline
\end{tabular}
\end{center}
\caption{Values of the dimension, setup, number of grid points, grid size, time step, time of duration of the simulation and modes for the runs that are used for the wave dispersion test. These 4 sets of parameters are associated to 8 runs, considering the 2 polarization, left and right in each cases.}
\label{table:dispersion}
\end{table}

In this section, we test whether the hybrid solver correctly captures the dispersion of waves propagating parallel to a uniform magnetic field
%supported by the plasma
in the hybrid formalism.
We initialize both 1D and 2D simulations, each containing an arbitrary set of initial modes (4 and 3 modes for 1D and 2D, respectively)
in a Maxwellian electron-proton plasma initially stationary with a density equal to $1$, a proton thermal velocity equal 0.01 and a electron thermal velocity equal to $0$.
For 1D runs, the initial background magnetic field is set to $B_x=1$ and circularly polarized perturbations are imposed on $B_y$ and $B_z$.
For 2D runs,
%it
the initial background magnetic field
is oriented along the diagonal of the domain, and the analysis is made on magnetic components perpendicular to this direction.
Two setups, hereafter called \textit{low} and \textit{high}, are ran in both 1D and 2D, to capture the wave dispersion at large and small scales, respectively.
In each final setup, an arbitrary number of wave modes $m$ are imposed initially.
%\rsm{In order to correctly capture the features of the waves, we have two contexts, namely "low" and "high", addressing the low and high values of the wave numbers $k$. For that purpose, the wave number is $k_m=2 \pi m/L_x$ in the 1D simulations
%($m$ being the mode indicated in the last row of Table~\ref{table:dispersion})
%, and $k_m = 2 \pi m / \sqrt{L_x + L_y}$ for the 2D simulations.
%in order to span the low and high ($\omega, k$) pairs with appropriate parameters.
The dimensionality, number of cells, mesh resolution, time step, simulation duration and excited modes numbers are summarized in Tab.\ref{table:dispersion} for all used setups.
In all simulations, the hyper resistivity coefficient $\nu$ is set to $0.005$, the initial amplitude of the excited modes is $0.01$ and their phase is random.
Each simulation is performed twice, with the excited mode imposed to have either Left- or Right-handed polarization. Then, the Fourier transform of $B_y \mp i B_z$ is computed, the sign depending on the wave polarization.
For each mode $m$ associated to the excited wave number $k_m$, the self-consistent frequency $\omega_m$ is found as a local maximum in spectral energy, which is then reported on Fig \ref{fig:wk}.
The theoretical fluid dispersion relations for the
%Alfvénic (eq. \ref{eq:wa}), and
Left-Alfvén-Ion-Cyclotron (eq. \ref{eq:wl}) and Right-Whistler (eq. \ref{eq:wr}) modes are also shown on the same figure with the analytical expressions:
%.

\begin{eqnarray}
   %\omega_A & = & k\label{eq:wa}\\
   \omega_L & = & \frac{k^2}{2}\left(\sqrt{1+4/k^2}-1\right)\label{eq:wl}\\
   \omega_R & = & \frac{k^2}{2}\left(\sqrt{1+4/k^2}+1\right)\label{eq:wr}
\end{eqnarray}

The Alfvén-Cyclotron and Whistler modes follow the fluid theory very well in such a cold plasma. 
%These computations are now integrated as functional tests so they are routinely executed at each modifications of the source code of the project.
After a first visual inspection of the results, the $\omega_m$ values are tabulated and the functional tests asserts the code continues finding them during each nightly build.

%The theoretical dispersion of the L mode also follows very well the simulation results at low frequencies as it diverges from the Alfvén mode. At higher frequencies the waves start to be affected by Landau damping which the theory does not take into account.

\subsubsection{Right-hand resonant streaming instability}

When a beam of charged particles is moving across a magnetized plasma at rest, three kinds of instabilities can develop \cite{gary1993}.
The resonant left-hand and right-hand modes, and the non-resonant mode.
There is a range of plasma parameters for which only one mode is unstable or where others have a much smaller growth rate\cite{gary1985}.
It is thus possible for us to select one of the resonant modes to assess the ability of the code to solve kinetic physics.
The right-hand resonant mode is chosen because it results in the most localised structures in phase space, allowing us to test their transport through patch level boundaries (see section \ref{sec:phasespacelevel}).
We use a setup already investigated \cite{gary1985} which consists of a main population of protons at rest, with a density equal to $1.0$, and a proton beam population, having a density of $0.01$ and a bulk velocity of $5.0$ embedded in a $B_x=1$ magnetic field.
These two populations are initially Maxwellian with a temperature equal 0.1. The electron temperature is also set to 0.1 so the most unstable mode is numerically predicted with a wavelength $\sim 33$ and a growth rate $\sim 0.09$.

\begin{figure}
    \centering
    \includegraphics[width=\linewidth]{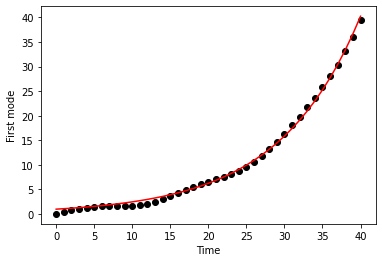}
    \caption{Time evolution of the first mode of the Fourier transform of $B_y - i B_z$ (black circles) and the associated exponential fit (red curves).}
    \label{fig:gamma}
\end{figure}

We used a computational domain of length 33 which is the wavelength of the most unstable mode. 
The simulation is performed with two fixed refinement levels to assess the validity of the multi-grid system as discussed in section \ref{sec:phasespacelevel}. 
We used 100 macroparticles per cell for the main population, and 1000 for the beam, in order to increase the statistical description of the resonant particles. The unstable mode being right-hand circularly polarized, we compute every $t=1$ the spatial Fourier transform of $B_y - i B_z$ and then reported with black solid circles in Fig.~\ref{fig:gamma} the modulus of the first mode, that is the one whose wavelength equals the length of the computational domain, the red curve is an exponential fit of these values.
%In this run with refinement, we obtain a growth rate $\gamma$ = 0.0932 which is in very good agreement with theoretical predictions.
%We performed similar runs with 100 particles per cell for the beam, with exact splitting, with a second order shape factor, with a third order shape factor and without refinement. We obtain a mean value for the growth rate of 0.0908, with a standard deviation of 0.0053. The linear phase of this mode ceases at $t\sim 45$.
The computational results from several simulations clearly show that, even with a magnetic seed for the magnetic perturbation of amplitude 0.1, the beginning of the linear phase and then the time at which the linear mode is saturating is varying from one run to another, as it depends on the initial microscopic state.
We thus have performed 200 runs for which we identify the saturation time, before which the growth of the instability is evaluated by a fit to an exponential.
We obtain an averaged growth rate of 0.09 with a standard deviation smaller than 1\% of this average, recovering the numerical values previously obtained \cite{gary1985}.

\subsection{Validation with mesh refinement}

Previous tests assessed the ability of the code to obtain solutions expected in the hybrid formalism.
In the following parts, we now test the behavior of the code with refined meshes, first refined statically from the initial condition, then dynamically by tagging regions of interest.

\subsubsection{Alfvén wave}
\begin{figure}
    \centering
    \includegraphics[width=\linewidth]{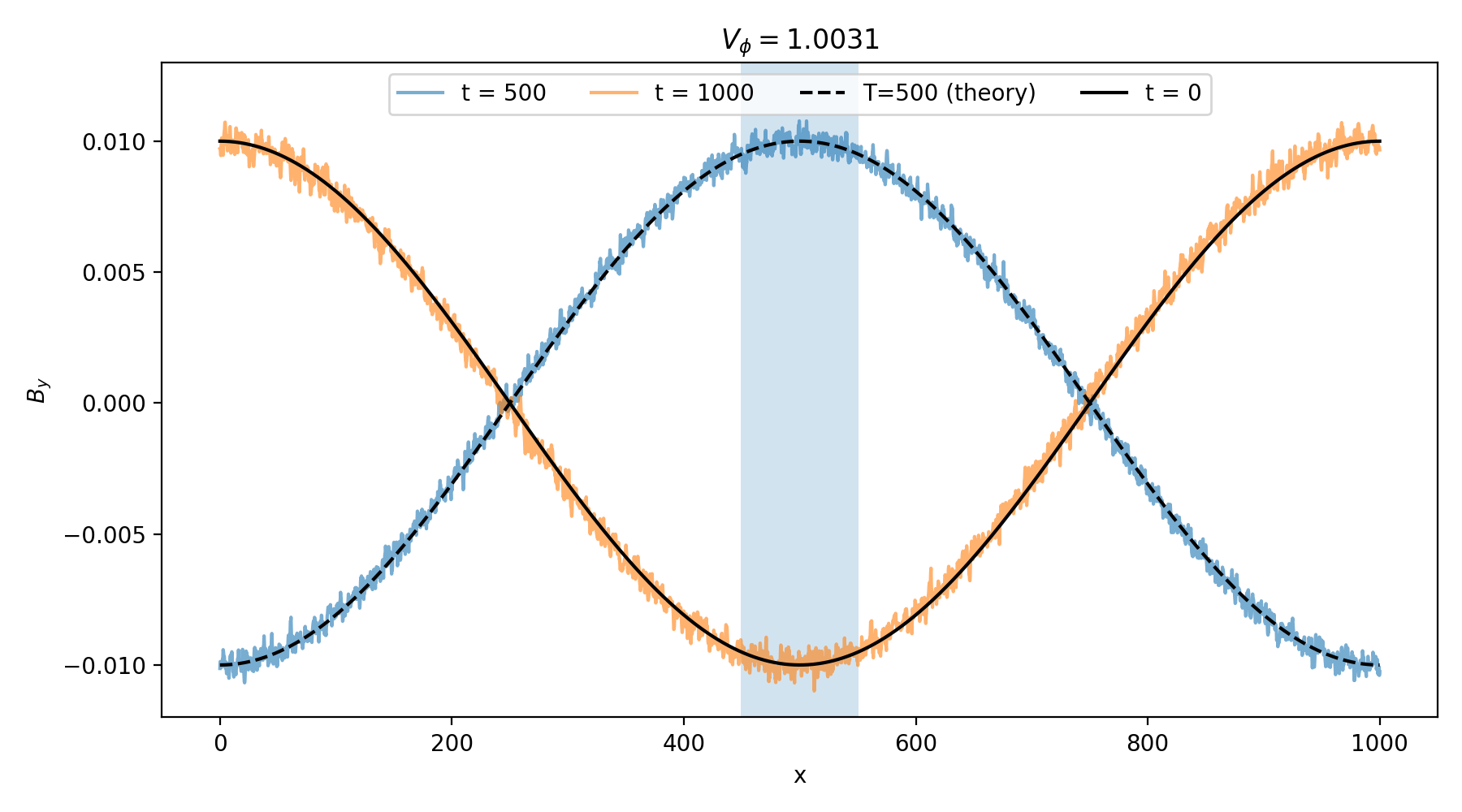}
    \caption{Alfvén wave moving across a fixed refined region of space. The solid black, blue and orange lines represent the $B_y$ component at times 0, 500 and 1000, respectively. The dashed black line represents $0.01\cos\left(2\pi/1000 x + 2\pi/1000t\right)$ for t=500. The blue area is the refined portion of the domain.}
    \label{fig:alfven}
\end{figure}

The goal of this test is to assess the distortion an Alfven wave may experience as it propagates across patch and level boundaries in 1D. We initialize \PHARE{} with an Alfvén wave of amplitude $\delta B_y=0.01$ and a wavelength
equal to the length of the computational domain.
%the size of the computational domain size. 
The whole domain at the coarsest level consists of 4000 cells for a total size of $L_x=1000$ and a time step $dt=0.001$. The domain is refined between $x=455$ and $x=550$ with only one refined level. %There are 80 patches of 50 cells each on the coarsest level and 15 patches of sizes between 50 and 98 on the refined level.
A single proton population with unit density and thermal velocity $v_{th}=0.01$ is used. The simulation is ran with 8 MPI processes until $t=1000$, corresponding to the time at which the wave should have traversed the whole domain. Figure \ref{fig:alfven} shows the $B_y$ component of the magnetic field at t=0, t=500 and t=1000. One can see the wave is exactly at the expected position at both t=500 and t=1000. A fit to $B_0\cos\left(kx+\phi(t)\right)$ is performed every $\Delta t_{\mathrm{fit}}=0.1$, to obtain the numerical values of $B_0$, $k$ and $\phi$, the amplitude, wave number and phase of the wave, respectively. This allows us to calculate the phase speed $V_\phi = \dot{\phi}/k$ as a function of time, which is found to be $V_\phi = 1.0031 \pm 0.0292$.
\new{The discrete jump in spatial resolution at the boundary of patch levels prevents waves at the smallest wavelength to exit fine regions. The accumulation of these waves at level boundaries can have dramatic impact regarding the solution.
Several codes employ filters\cite{fujimoto2011}, or absorbing layers \cite{2004CoPhC.164..171V} to deal with these issues.
In our tests, we have sometimes observed spurious accumulation of small wavelength waves in cases where the refinement is fixed in time and when there is a fast flow normal to the level boundary, convecting and accumulating waves upon it.
Generally, the accumulation takes many cyclotron times to be noticeable and disappears when hyper-resistivity is increased, introducing a faster dissipation time scale.
When the refinement is dynamical, howvever, no accumulation is seen, which is the result of the frequent motion of the level boundaries which prevents wave accumulation.
}
Here, after a million time steps, no noticeable artifact appears on the signal although it has crossed many patch boundaries and the two level boundaries. This tests is automatically executed each time the code is updated and set to fail if the absolute error on the calculated phase velocity is larger than $0.05$.

\subsubsection{Moving tangential discontinuity}\label{sec:movingtd}
In this test we setup two sharp tangential discontinuities in a globally rightward moving plasma with a velocity $V_x=2$.
In a fluid framework, one expects the two tangential discontinuities to be simply advected at the prescribed velocity and to keep a profile identical to the initial one in the co-moving frame.
In this kinetic framework however, the discontinuity is not a kinetic equilibrium and thus can have its structure slightly altered within the few first ion cyclotron times. Finite diffusion can also decrease its sharpness over time.
Two simulations are ran, one with adaptive mesh refinement and one without. The non-refined simulation has a mesh size equal to the coarsest mesh size of the refined simulation, set to $\Delta x=1$.
It is enough in regions where the plasma and field are homogeneous, but poor in the discontinuity regions since it equals their half-thickness. 
The refined run is allowed to create up to 3 levels, i.e. to refine the mesh up to $\Delta x=0.25$.
The goal of this test is twofold :
%.
%The first goal is to
(i) ensure that in the refined case, \PHARE{} refines the solution around the location of the discontinuities at all times
%.
and (ii)
%The second objective is to
compare the evolution of the structure of the discontinuity between the refined and non-refined runs.
The domain extends over $L_x=200$.
The initial density is $n=1$, the magnetic field is given by $B_z=0.5$ and $B_y = -1 + 2\left(S\left(x, 0.25L_x,1\right) - S\left(x, 0.75L_x,1\right)\right)$ where

\begin{equation}\label{eq:S}
     S\left(y,y^\star,\lambda\right) = 0.5\left(1 + \tanh\left(\frac{y-y^\star}{\lambda}\right)\right)
\end{equation}

\begin{figure}
    \centering
    \includegraphics[width=\linewidth]{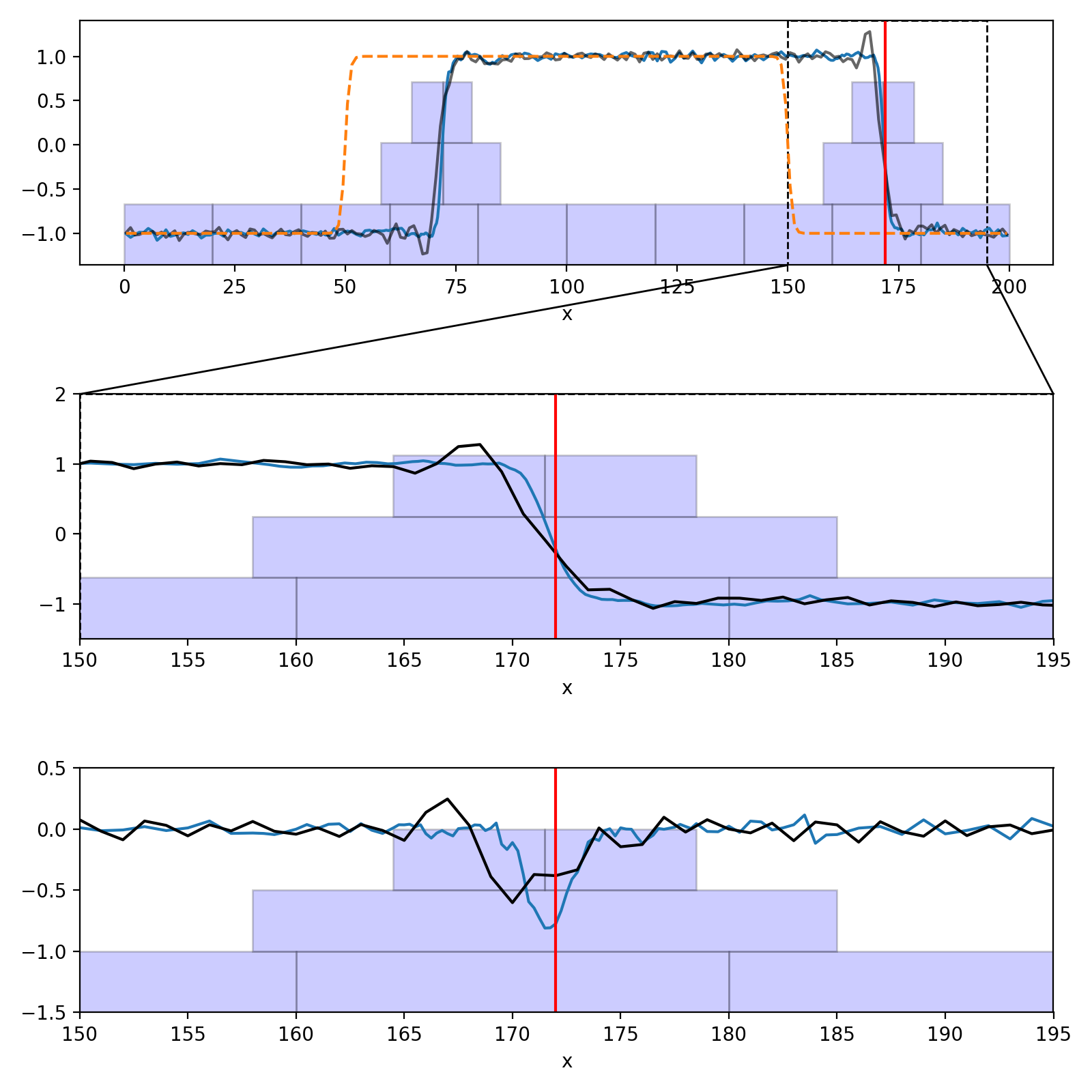}
    \caption{Top panel shows $B_y$ for both runs for the whole domain. The orange dashed-line represent the initial profile. Middle panels compares of the $B_y$ component of the magnetic field at $t=11$ without refinement (black line) and with refinement (blue line). Bottom panel has the same format for the $J_z$ component of the current density. The mesh size of the non-refined simulation is $dx=1$, which is equal to the lowest resolution of the refined simulation. Intermediary and finest resolutions have $dx=0.5$ and $dx=0.25$, respectively. The vertical red line at $x=172$ marks the theoretical center of the discontinuity at that time. In all panels, light blue rectangles represent the extent of patches, whose refinement level increases vertically. }
    \label{fig:TD}
\end{figure}

This produces two discontinuities centered at $x_0=50$ and $x_1=150$, respectively, of half-width $1$.
A single proton population is initialized with 100 particles per cell with first order shape function.
The ion temperature is chosen to satisfy at each point the total pressure balance: $T(x) = 1 - 0.5B(x)^2$.
The hyper-resistivity $\nu$ is set to  $0.01$, the electron temperature is set to 0.
The total simulated time is $T=20$ with $2000$ time steps.
 Figure \ref{fig:TD} shows the magnetic field component $B_y$ and the associated current density component $J_z$ along $x$ at $t=11$ for both runs, in a zoom
%in
of the region of the rightmost discontinuity.
Given its initial position, and the bulk flow, the discontinuity should in theory be centered at $x=172$ at this time.
One can first notice that \PHARE{} has created 3 patch levels with two refined ones.
Each refined level is composed of two patches and is centered in the region of the discontinuity, as expected.
Then, one can notice the discontinuity has moved in both runs and roughly arrives at the same location.
However if the peak current density is located exactly where expected in the refined solution, it is not the case in the other run.
The non-refined run sees its discontinuity profile altered by spurious downwind oscillations and a somewhat decreased slope.
This alteration results in a current density which peak value is reduced by half compared to its initial value and its global shape is different.
This test is also automatically ran each time the code is updated, it fits the final position of the discontinuity and asserts its width and position are consistent with their initial values.

\subsubsection{Continuity of phase space across level boundaries}\label{sec:phasespacelevel}
We previously discussed how the code recovers the growth of the right hand resonant mode associated to the ion streaming instability. 
This instability will lead beam particles to give their energy to electromagnetic fluctuations, structuring the phase space density in the process.
Our goal in this test is to assess how well these self-consistent phase space density structures are transported across (fixed) level boundaries.

%\todo{Plot the phase space of the 2stream and discuss that structures are continuous across patch (as expected) and level (as hoped) boundaries.}

\begin{figure}
    \centering
    \includegraphics[width=\linewidth]{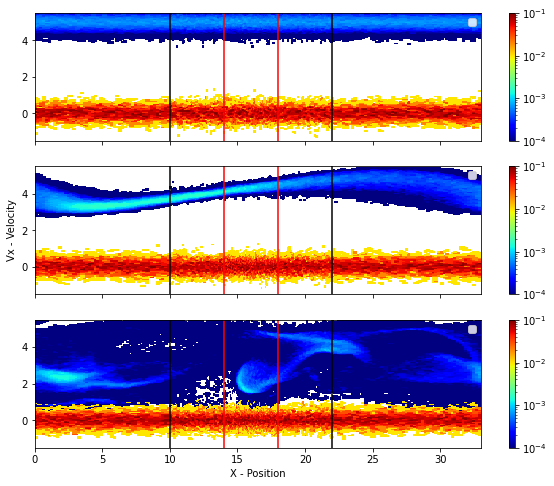}
    \caption{logarithmic value of the distribution function depending on the $X$ position and $V_X$ velocity for main and beam population at $t=0$ (upper panel), $t=64$ (middle panel) and $t=85.4$ (lower panel). The two vertical black lines are framing the first level of refinement and the two vertical red lines the second one.}
    \label{fig:dist}
\end{figure}

In Fig.~\ref{fig:dist}, the particle density in phase space $(x,V_x)$ is displayed at $t=0$ (upper panel), $t=64$ (middle panel) and $t=85.4$ (lower panel) for the same two-stream run that was previously discussed.
The upper panel depicts the initial setup.
The middle panel shows the beginning of the non-linear phase of the right hand resonant mode where the beam is corrugated with the $m=1$ mode clearly dominating.
We can also notice the decrease of the bulk velocity of the beam, as its associated energy is efficiently converted to magnetic energy.
The time in the lower panel has been chosen to emphasize that these complex, non-linear phase space structures cross the level boundaries without any alteration. 
%\rsm{This test outlines that the physics at play between the different levels is the same, meaning that the refinement or coarsening  operations between the levels has no consequences on the physical growth of the instability.}
%This highlights the very good behavior of the optimized splitting strategy\cite{Smets:2021dw} that we used in this numerical simulation, and shows no clear differences from a statistical point of view with the exact splitting case already reported above.

\subsubsection{Magnetic reconnection}

We now show results in two dimensions, with simulations of the magnetic reconnection process.
Magnetic reconnection occurs in magnetized plasmas where the magnetic field changes its orientation over a short distance, forming a current sheet.
Reconnection is a self-driven process where magnetic tension from the newly connected field lines expels the plasma from the reconnection site, thereby pulling in upstream magnetic flux and the frozen in plasma.
Here we setup a double periodic system where current sheets are positioned at $y_1=0.3L_y$ and $y_2=0.7L_y$, in a domain box that is $L_x=L_y=40$ wide. The magnetic field profile is given by

\begin{equation}\label{eq:harrisB}
    B_x(y) = B_1 + \left(B_2 - B_1\right)\left(S\left(y,y_1,0.5\right) - S\left(y,y_2,0.5\right)\right)
\end{equation}

\noindent
and $B_1=-1$, $B_2= 1$ a,s $S$ is given by eq. \ref{eq:S}.
A small magnetic perturbation is superimposed to this magnetic field at the center of the two current sheets.
The electron temperature is set to $0$.
There is only one ion population made of protons, which particle density is given by

\begin{equation}\label{eq:harrisN}
    n(y) = n_b + \cosh^{-2}\left(\frac{y-y_1}{\lambda}\right) + \cosh^{-2}\left(\frac{y-y_2}{\lambda}\right)
\end{equation}

\noindent
where the background density is $n_b=0.4$. The ion temperature is chosen so that the total pressure $K = 0.7$ is initially uniform in the whole domain, and is given by
%is balanced

\begin{equation}\label{eq:harrisT}
    T(y) = \frac{1}{n}\left(K - 0.5B^2(y)\right)
\end{equation}

\noindent
%with the total pressure $K = 0.7$.
%The electron temperature is set to $0$.

%\onecolumn
\begin{figure*}
    %\centering
    \includegraphics[width=\textwidth]{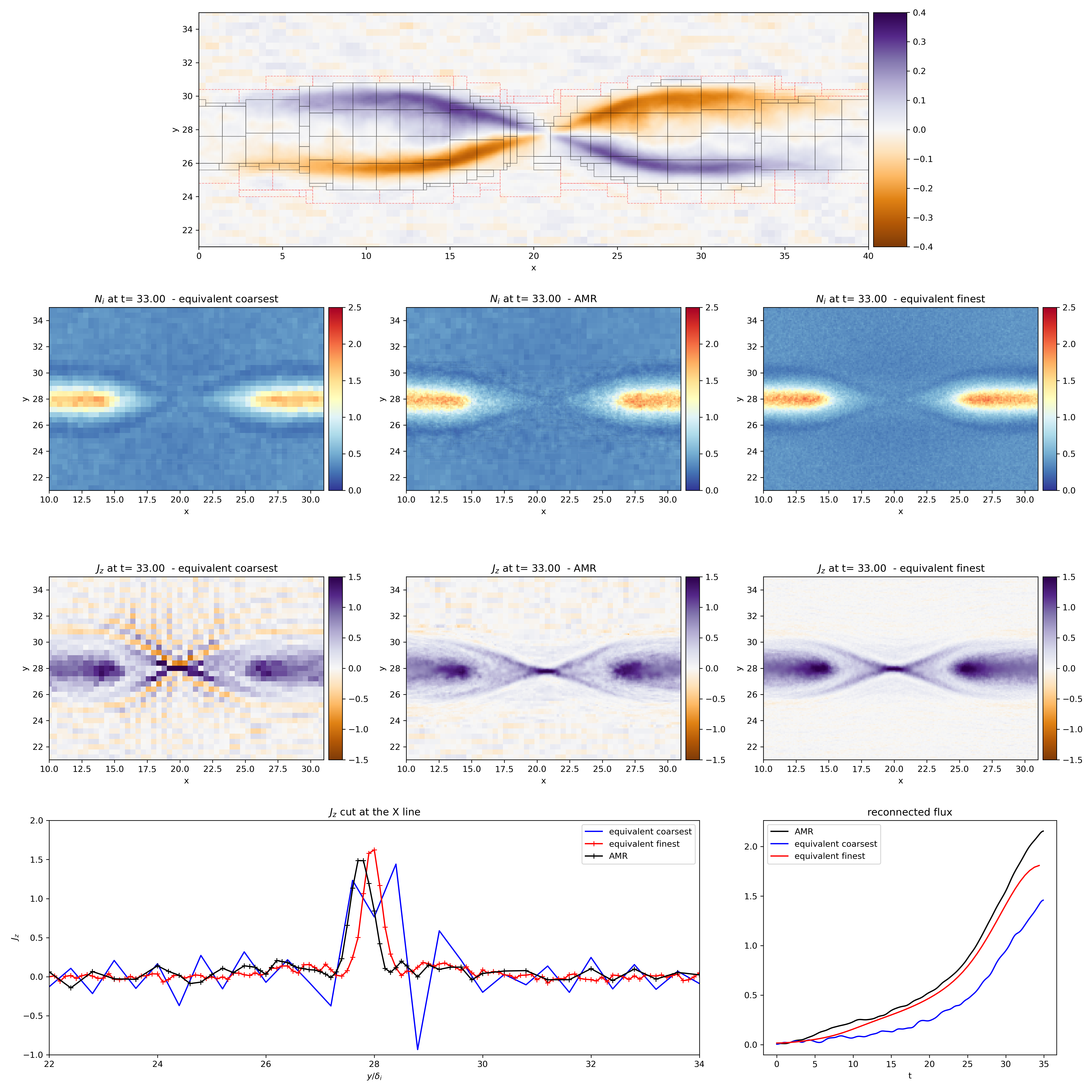}
    \caption{Top panel shows the out of plane component $B_z$ color coded. Red and black rectangles represent the borders of patches of the $L_1$ and $L_2$ refinement levels, respectively. The panels on the second and third row represent, at the same time $t=3$, the total ion density and the out of plane current density, respectively. On these two rows, the middle panel represent the result of the AMR simulation with 3 mesh levels. The leftmost panel represent the result of a uniform mesh run at the coarsest resolution of the AMR one, so-called \textit{equivalent coarsest} run. The rightmost panel is the so-called \textit{equivalent finest} run, with a uniform mesh at the finest resolution of the AMR run. \new{The leftmost panel on the bottom row shows a cut of $J_z$ along $y$ through the X line. The rightmost panel shows the time evolution of the reconnected flux for each of the three runs.}}
    \label{fig:reconnection}
\end{figure*}
%\onecolumn

Three simulations are performed. A first simulation, with AMR enabled, is ran with a coarsest spatial resolution set to $\Delta x=\Delta y=0.4$ and the associated time step to $\Delta t=0.005$.
The maximum number of levels is set to 3, which means 2 refined levels.
A second simulation is performed, so called \textit{equivalent coarsest}, with a uniform mesh at the coarsest resolution of the AMR run and all other parameters identical.
The third simulation, called the \textit{equivalent finest}, with the same parameters as the other two except now the spatial resolution is $dx=dy=0.1$ like the finest one allowed in the AMR run.
In all runs the resistivity is set to $\eta=0.001$ and hyper-resistivity to $\nu=0.002$.

Figure \ref{fig:reconnection} shows the result of all simulations, zoomed in one of the two reconnection sites.
The top panel shows the out of plane magnetic field component $B_z$, on top of which the borders of $L_1$ and $L_2$ patches are shown.
The magnetic field reveals the well-known quadrupolar pattern associated with the Hall effect \cite{aunai2013dissip}.
$L_1$ patches are located in a narrow band between $y=24$ and $y=31$, and $L_2$ patches are nested in that band, mapping even closer the gradients of the magnetic field structure.
The three panels below show the ion density at the same time.
All simulations show the expected density pile-up in the downstream $X$-direction resulting from the expulsion of the initial current sheet with high density plasma downstream towards the periodic boundaries.
In all simulations, the density is lower on the separatrices, a typical feature of collisionless reconnection\cite{1998JGR...103.9165S}.
Panels in the third row show the out of plane current density.
The current density is greatly enhanced around the reconnection site to sustain the large magnetic reversal over the sub-ion scale distance.
It is also enhanced on regions connected to the reconnection site, i.e. separatrices.
On the left and right side of the reconnection site, the current density increases and forms a broader sheet that results from the pile-up of reconnected magnetic flux against the initial tangential sheet and the periodic boundary.
\new{The figure also shows the out of plane component $J_z$ of the current density along $y$ in a cut through the X line, for each run.}
The finest uniform simulation clearly resolves the hyper-resistive current sheet \cite{aunai2013dissip}.
The coarsest simulation, however, sees the current sheet collapse down to the grid scale, resulting in many numerical artifacts around the reconnection site.
The AMR simulation in the middle panels shows the current sheet is well resolved as in the equivalent finest run.
\new{This can also be seen in the cuts along $y$, for which the current layer is resolved by many more points in the AMR and equivalent finest runs, whereas the structure of the layer is strongly altered and thicker in the equivalent coarsest case.}
In the AMR run, the spatial resolution quickly becomes coarse upstream of the reconnection site, however and contrary to the equivalent coarsest run, no numerical artifact is observed.
\new{The time evolution of the reconnected flux for each run is also reported on Fig. \ref{fig:reconnection}. Clearly the AMR run is much closer to the reference equivalent finest run than the equivalent coarsest is. The coarse run reaches a similar (somewhat smaller) reconnection rate in the fast phase for times later than $t\sim 20$, and most of the difference in the amount of flux that is ultimately reconnected stands in how reconnection struggles a bit more to start in that coarsest version. The similarity in the reconnection rate in the fast phase is not very surprising as it is known that for symmetric conditions and for such a small domain, the rate is quite insensitive to the mechanism that breaks the field line connectivity (here numerical errors). In asymmetric conditions, and for larger domains, however, the lack of a proper dissipation mechanism is known to potentially lead to a drastically different macroscopic evolution, possibly dominated by spurious current sheet elongations and plasmoid formation\cite{aunai2013dissip}.
Preventing artifacts and providing a proper dissipation in the coarsest resolution run would require a significant increase of the the hyper-resistivity, with the drawback that it would also start to dissipate the magnetic field at ion scales and alter the dynamics of the reconnection process under study.
The cuts of the current density also show that the AMR peak is slightly shifted on the left of that of the one from the equivalent finest run.
This is due to the full periodicity of the domain where the two layers located at $0.3L_y$ and $0.7L_y$, and to the somewhat sooner start of reconnection in the AMR run.
At the same time, the AMR run has reconnected slightly more flux than the equivalent finest one. The location of the layer means more flux is processed between $0.3L_y$ and $0.7L_y$ than "outside", thereby bringing the AMR layers closer to each other than in the equivalent finest run. Consistently the lower layer is seen (not shown) slightly shifted to the right with respect to that of equivalent finest one.
}
These results confirm that the mesh refinement in \PHARE{} is able to map and resolve fine scale structures at high resolution.

\section{Software details}\label{sec:software}

Most of the \PHARE{} code base is written in C++. The high level user input/output interface is written in Python.
The code is developed in an open source fashion\footnote{https://github.com/PHAREHUB/PHARE}.
The management of adaptive mesh refinement instrinsic details is handle by the massively parallel library SAMRAI \cite{gunney16}.
SAMRAI performs operations related to the definition of the position and geometry of the patches, through a tile clustering algorithm\cite{gunney16}.
%It also partitions the patches among the different MPI processes, here the so-called cascade partitioner, also provided by SAMRAI.
%, is being used.
\new{In PIC codes, the spatial distribution of macroparticles in the domain introduces a difficulty to balance the work load across the processes.
In \PHARE{}, patches are distributed among the different MPI processes.
The partitioner algorithm is provided by the SAMRAI library.
The patches are currently being distributed based on the number of computational cells they count.
Since macroparticles are initially uniformly distributed in the domain (their statistical weight varies with the density), this strategy balances the initial load.
But the self-consistent evolution of the plasma will further create load imbalance.
This strategy is thus considered as a first step and work is currently performed to use the so-called \emph{cascade partitioner} implemented in SAMRAI \cite{gunney16}, that can account for an inhomogeneous load within patches by counting the number of particles in each cell.
Work is also currently being performed to add another level of parallelisation to further decompose the work load within MPI processes across CPU and GPU threads.
This paper being dedicated to the validity of the AMR methodology, the code is not fully optimized yet and an exhaustive performance analysis will be the topic of a future communication.
}
%The scaling of SAMRAI AMR components have recently been shown to behave well up to 1.5M cores and 2M MPI processes.

SAMRAI manipulates abstractions of the data defined on the patches, its transfers and copies, and the determination of geometrical overlaps between patches of the same level or of different ones.
Concrete implementation of these data, transfers and geometrical operations are developed in \PHARE{} and follow the concepts explained in above sections.
The code is written in such a way that components solving the hybrid equations and those transferring information to the borders of refinement levels, and coarsening data to the next coarser level, are hidden behind abstractions.
In this way, despite that here only hybrid PIC equations and data transfers are implemented, the code architecture is already made to support other possible implementations.
Future work will in particular focus on implementing a fluid hybrid model, operating on the coarsest levels of the patch hierarchy.
The SAMRAI library has already been used in such a multi-formalism simulations \cite{2002APS..DFD.KA002W}.
In this way, \PHARE{} could model large fluid domains in which hybrid PIC patches of incrementally fine resolution are created dynamically over critical regions.

\section{Summary and perspective}\label{sec:sumpersp}
This paper presents the code \PHARE{}, evolving hybrid PIC equations using a patch based AMR approach.
Hybrid equations are solved on uniform resolution cartesian patches with a predictor-predictor-corrector temporal scheme and a Yee spatial discretization.
Macroparticles are modeled with B-spline interpolation kernels of order 1, 2 or 3 and their trajectory is calculated with the Boris algorithm.
Any number of ion populations can be modeled.
The AMR methodology adopted here is inspired from the MLMD method where all refinement levels evolve their fields and their macroparticle populations, which allows not to depend on macroparticle merging.
The evolution of the patch hierarchy is done through a recursive procedure using a time step divided by 4 between each refinement level, where the spatial resolution is increased by a factor of 2.
Validation tests have been presented.
They reveal the code successfully captures the expected dispersion of waves in the hybrid kinetic regime.
The solver has also been demonstrated to accurately predict the growth rate of the kinetic ion streaming instability.
Structures in phase space developing during the nonlinear phase of the instability have been shown to propagate through level borders without alteration. A large scale Alfvén wave was shown to propagate in a system with two grid levels without alteration at level borders and with an accurate phase velocity.
The adaptive meshing has first been demonstrated in 1D to be able to capture two tangential discontinuities advected in a global flow.
The refinement was demonstrated to be able to conserve the initial magnetic field profile whereas coarser simulation cannot, resulting in distortions in the current sheet structure.
Finally, the code is validated in a 2D system and shown to capture well-established features of collisionless magnetic reconnection.
The adaptive meshing adequately maps the structures of the magnetic field and helps resolving the sub-ion scales without needing a uniform high resolution mesh.

This paper validates the method behind the making of a patch-based AMR hybrid PIC code.
Future work will now focus on the several following points.
First, extend the code to simulate three-dimensional systems.
All the code components are already implemented in 3D. 
Remaining work however remains in its validation and the development of 3D post-processing tooling.
3D runs also call for optimizations of the various components of the code.
The code benefits from distributed parallelization through MPI. Multithreading and GPU parallelizations are not yet finished and important for large 3D runs.
Work is also being done to design a multi-formalisms patch hierarchy where the coarsest levels would evolve a fluid hybrid system coupled to dynamically created hybrid kinetic refined levels.

\section*{Acknowledgements}
The authors acknowledge support from the plasapar labex and federation, from the Laboratory of Plasma Physics, from the Centre National d'Etudes Spatiales (CNES) and Programme National Soleil-Terre (PNST). We also acknowledge the SAMRAI developer team for their help. \new{The authors acknowledge the two reviewers for their questions and comments that helped us improving the quality of the manuscript.}

\bibliography{phare1d}

\begin{thebibliography}{10}
\expandafter\ifx\csname url\endcsname\relax
  \def\url#1{\texttt{#1}}\fi
\expandafter\ifx\csname urlprefix\endcsname\relax\def\urlprefix{URL }\fi
\expandafter\ifx\csname href\endcsname\relax
  \def\href#1#2{#2} \def\path#1{#1}\fi

\bibitem{Dargent2020}
J.~Dargent, N.~Aunai, B.~Lavraud, S.~Toledo-Redondo, F.~Califano,
  \href{http://adsabs.harvard.edu/cgi-bin/nph-data\_query?bibcode=2020GeoRL..4786546D\&link\_type=EJOURNAL}{{Simulation
  of Plasmaspheric Plume Impact on Dayside Magnetic Reconnection}}, Geophysical
  Research Letters 47~(4) (2020) e86546.
\newblock \href {https://doi.org/10.1029/2019gl086546}
  {\path{doi:10.1029/2019gl086546}}.
\newline\urlprefix\url{http://adsabs.harvard.edu/cgi-bin/nph-data\_query?bibcode=2020GeoRL..4786546D\&link\_type=EJOURNAL}

\bibitem{DaughtonNat2011}
W.~Daughton, V.~Roytershteyn, H.~Karimabadi, L.~Yin, B.~Albright, B.~Bergen,
  K.~Bowers,
  \href{http://adsabs.harvard.edu/cgi-bin/nph-data\_query?bibcode=2011NatPh...7..539D\&link\_type=ABSTRACT}{{Role
  of electron physics in the development of turbulent magnetic reconnection in
  collisionless plasmas}}, Nature Physics 7~(7) (2011) 539 -- 542.
\newblock \href {https://doi.org/10.1038/nphys1965}
  {\path{doi:10.1038/nphys1965}}.
\newline\urlprefix\url{http://adsabs.harvard.edu/cgi-bin/nph-data\_query?bibcode=2011NatPh...7..539D\&link\_type=ABSTRACT}

\bibitem{Filippychev2002}
D.~S. Filippychev, {Hybrid Simulation of Space Plasmas: Models with Massless
  Fluid Representation of Electrons. V. Reconnection of Magnetic Field Lines},
  Computational Mathematics and Modeling 13~(3) (2002) 215--248.
\newblock \href {https://doi.org/10.1023/a:1016003831330}
  {\path{doi:10.1023/a:1016003831330}}.

\bibitem{lipatov2002}
A.~S. Lipatov, {The Hybrid Multiscale Simulation Technology, An Introduction
  with Application to Astrophysical and Laboratory Plasmas}, Scientific
  Computation, 2002.
\newblock \href {https://doi.org/10.1007/978-3-662-05012-5}
  {\path{doi:10.1007/978-3-662-05012-5}}.

\bibitem{2016JCoPh.309..295L}
L.~Leclercq, R.~Modolo, F.~Leblanc, S.~Hess, M.~Mancini,
  \href{http://adsabs.harvard.edu/cgi-bin/nph-data\_query?bibcode=2016JCoPh.309..295L\&link\_type=EJOURNAL}{{3D
  magnetospheric parallel hybrid multi-grid method applied to planet-plasma
  interactions}}, Journal of Computational Physics 309 (2016) 295 -- 313.
\newblock \href {https://doi.org/10.1016/j.jcp.2016.01.005}
  {\path{doi:10.1016/j.jcp.2016.01.005}}.
\newline\urlprefix\url{http://adsabs.harvard.edu/cgi-bin/nph-data\_query?bibcode=2016JCoPh.309..295L\&link\_type=EJOURNAL}

\bibitem{2017JGRA..122.2877H}
S.~Hoilijoki, U.~Ganse, Y.~Pfau-Kempf, P.~A. Cassak, B.~M. Walsh, H.~Hietala,
  S.~v. Alfthan, M.~Palmroth,
  \href{http://adsabs.harvard.edu/cgi-bin/nph-data\_query?bibcode=2017JGRA..122.2877H\&link\_type=EJOURNAL}{{Reconnection
  rates and X line motion at the magnetopause: Global 2D-3V hybrid-Vlasov
  simulation results}}, Journal of Geophysical Research (Space Physics) 122~(3)
  (2017) 2877 -- 2888.
\newblock \href {https://doi.org/10.1002/2016ja023709}
  {\path{doi:10.1002/2016ja023709}}.
\newline\urlprefix\url{http://adsabs.harvard.edu/cgi-bin/nph-data\_query?bibcode=2017JGRA..122.2877H\&link\_type=EJOURNAL}

\bibitem{guo2021}
J.~Guo, S.~Lu, Q.~Lu, Y.~Lin, X.~Wang, Q.~Zhang, Z.~Xing, K.~Huang, R.~Wang,
  S.~Wang, {Three‐Dimensional Global Hybrid Simulations of High Latitude
  Magnetopause Reconnection and Flux Ropes During the Northward IMF},
  Geophysical Research Letters 48~(21) (2021).
\newblock \href {https://doi.org/10.1029/2021gl095003}
  {\path{doi:10.1029/2021gl095003}}.

\bibitem{guo2020}
Z.~Guo, Y.~Lin, X.~Wang, S.~K. Vines, S.~H. Lee, Y.~Chen, {Magnetopause
  Reconnection as Influenced by the Dipole Tilt Under Southward IMF Conditions:
  Hybrid Simulation and MMS Observation}, Journal of Geophysical Research:
  Space Physics 125~(9) (2020).
\newblock \href {https://doi.org/10.1029/2020ja027795}
  {\path{doi:10.1029/2020ja027795}}.

\bibitem{2013PhPl...20d2901A}
N.~Aunai, M.~Hesse, C.~Black, R.~Evans, M.~Kuznetsova,
  \href{http://adsabs.harvard.edu/cgi-bin/nph-data\_query?bibcode=2013PhPl...20d2901A\&link\_type=EJOURNAL}{{Influence
  of the dissipation mechanism on collisionless magnetic reconnection in
  symmetric and asymmetric current layers}}, Phys Plasmas 20~(4) (2013) 042901,
  accepted in Physics of Plasmas, 12 pages.
\newblock \href {https://doi.org/10.1063/1.4795727}
  {\path{doi:10.1063/1.4795727}}.
\newline\urlprefix\url{http://adsabs.harvard.edu/cgi-bin/nph-data\_query?bibcode=2013PhPl...20d2901A\&link\_type=EJOURNAL}

\bibitem{Palmroth2018Review}
M.~Palmroth, U.~Ganse, Y.~Pfau-Kempf, M.~Battarbee, L.~Turc, T.~Brito,
  M.~Grandin, S.~Hoilijoki, A.~Sandroos, S.~v. Alfthan,
  \href{https://doi.org/10.1007/s41115-018-0003-2}{{Vlasov methods in space
  physics and astrophysics}}, Living Reviews in Computational Astrophysics
  (2018) 1 -- 54\href {https://doi.org/10.1007/s41115-018-0003-2}
  {\path{doi:10.1007/s41115-018-0003-2}}.
\newline\urlprefix\url{https://doi.org/10.1007/s41115-018-0003-2}

\bibitem{lapenta2012sw}
G.~Lapenta,
  \href{http://adsabs.harvard.edu/cgi-bin/nph-data\_query?bibcode=2012JCoPh.231..795L\&link\_type=EJOURNAL}{{Particle
  simulations of space weather}}, Journal of Computational Physics 231~(3)
  (2012) 795 -- 821.
\newblock \href {https://doi.org/10.1016/j.jcp.2011.03.035}
  {\path{doi:10.1016/j.jcp.2011.03.035}}.
\newline\urlprefix\url{http://adsabs.harvard.edu/cgi-bin/nph-data\_query?bibcode=2012JCoPh.231..795L\&link\_type=EJOURNAL}

\bibitem{karimabadi2006tz}
H.~Karimabadi, H.~Vu, D.~Krauss-Varban, Y.~Omelchenko,
  \href{http://adsabs.harvard.edu/cgi-bin/nph-data\_query?bibcode=2006ASPC..359..257K\&link\_type=EJOURNAL}{{Global
  Hybrid Simulations of the Earth's Magnetosphere}}, Numerical Modeling of
  Space Plasma Flows: Astronum-2006 ASP Conference Series 359 (2006) 257.
\newline\urlprefix\url{http://adsabs.harvard.edu/cgi-bin/nph-data\_query?bibcode=2006ASPC..359..257K\&link\_type=EJOURNAL}

\bibitem{omelchenko2006DES}
Y.~A. Omelchenko, H.~Karimabadi,
  \href{http://adsabs.harvard.edu/cgi-bin/nph-data\_query?bibcode=2006JCoPh.216..153O\&link\_type=EJOURNAL}{{Event-driven,
  hybrid particle-in-cell simulation: A new paradigm for multi-scale plasma
  modeling}}, Journal of Computational Physics 216~(1) (2006) 153 -- 178.
\newblock \href {https://doi.org/10.1016/j.jcp.2005.11.029}
  {\path{doi:10.1016/j.jcp.2005.11.029}}.
\newline\urlprefix\url{http://adsabs.harvard.edu/cgi-bin/nph-data\_query?bibcode=2006JCoPh.216..153O\&link\_type=EJOURNAL}

\bibitem{Karimabadi:2007uv}
H.~Karimabadi, {A new methodology for multi-scale simulation of plasmas},
  advanced methods for space simulations (2007) 1 -- 9.

\bibitem{omelchenkohypers2012}
Y.~A. Omelchenko, H.~Karimabadi,
  \href{http://adsabs.harvard.edu/cgi-bin/nph-data\_query?bibcode=2012JCoPh.231.1766O\&link\_type=EJOURNAL}{{HYPERS:
  A unidimensional asynchronous framework for multiscale hybrid simulations}},
  Journal of Computational Physics 231~(4) (2012) 1766 -- 1780.
\newblock \href {https://doi.org/10.1016/j.jcp.2011.11.004}
  {\path{doi:10.1016/j.jcp.2011.11.004}}.
\newline\urlprefix\url{http://adsabs.harvard.edu/cgi-bin/nph-data\_query?bibcode=2012JCoPh.231.1766O\&link\_type=EJOURNAL}

\bibitem{JLVay2002}
J.~L. Vay, P.~Colella, P.~McCorquodale, B.~v. Straalen, A.~Friedman, D.~P.
  Grote,
  \href{http://adsabs.harvard.edu/cgi-bin/nph-data\_query?bibcode=2002LPB....20..569V\&link\_type=EJOURNAL}{{Mesh
  refinement for particle-in-cell plasma simulations: Applications to and
  benefits for heavy ion fusion}}, Laser and Particle Beams 20~(4) (2002) 569
  -- 575.
\newblock \href {https://doi.org/10.1017/s0263034602204139}
  {\path{doi:10.1017/s0263034602204139}}.
\newline\urlprefix\url{http://adsabs.harvard.edu/cgi-bin/nph-data\_query?bibcode=2002LPB....20..569V\&link\_type=EJOURNAL}

\bibitem{JLVay2004}
J.~L. Vay, P.~Colella, A.~Friedman, D.~P. Grote, P.~McCorquodale, D.~B.
  Serafini,
  \href{http://adsabs.harvard.edu/cgi-bin/nph-data\_query?bibcode=2004CoPhC.164..297V\&link\_type=EJOURNAL}{{Implementations
  of mesh refinement schemes for Particle-In-Cell plasma simulations}},
  Computer Physics Communications 164~(1) (2004) 297 -- 305.
\newblock \href {https://doi.org/10.1016/j.cpc.2004.06.075}
  {\path{doi:10.1016/j.cpc.2004.06.075}}.
\newline\urlprefix\url{http://adsabs.harvard.edu/cgi-bin/nph-data\_query?bibcode=2004CoPhC.164..297V\&link\_type=EJOURNAL}

\bibitem{fujimoto2006}
K.~Fujimoto, S.~Machida,
  \href{http://adsabs.harvard.edu/cgi-bin/nph-data\_query?bibcode=2006JCoPh.214..550F\&link\_type=EJOURNAL}{{Electromagnetic
  full particle code with adaptive mesh refinement technique: Application to
  the current sheet evolution}}, Journal of Computational Physics 214~(2)
  (2006) 550 -- 566.
\newblock \href {https://doi.org/10.1016/j.jcp.2005.10.003}
  {\path{doi:10.1016/j.jcp.2005.10.003}}.
\newline\urlprefix\url{http://adsabs.harvard.edu/cgi-bin/nph-data\_query?bibcode=2006JCoPh.214..550F\&link\_type=EJOURNAL}

\bibitem{Fujimoto:2008jv}
K.~Fujimoto, R.~D. Sydora,
  \href{http://linkinghub.elsevier.com/retrieve/pii/S0010465508000908}{{Electromagnetic
  particle-in-cell simulations on magnetic reconnection with adaptive mesh
  refinement}}, Computer Physics Communications 178~(12) (2008) 915 -- 923.
\newblock \href {https://doi.org/10.1016/j.cpc.2008.02.010}
  {\path{doi:10.1016/j.cpc.2008.02.010}}.
\newline\urlprefix\url{http://linkinghub.elsevier.com/retrieve/pii/S0010465508000908}

\bibitem{colella2010}
P.~Colella, P.~C. Norgaard,
  \href{http://adsabs.harvard.edu/cgi-bin/nph-data\_query?bibcode=2010JCoPh.229..947C\&link\_type=EJOURNAL}{{Controlling
  self-force errors at refinement boundaries for AMR-PIC}}, Journal of
  Computational Physics 229~(4) (2010) 947 -- 957.
\newblock \href {https://doi.org/10.1016/j.jcp.2009.07.004}
  {\path{doi:10.1016/j.jcp.2009.07.004}}.
\newline\urlprefix\url{http://adsabs.harvard.edu/cgi-bin/nph-data\_query?bibcode=2010JCoPh.229..947C\&link\_type=EJOURNAL}

\bibitem{muller2011}
J.~Müller, S.~Simon, U.~Motschmann, J.~Schüle, K.-H. Glassmeier, G.~J.
  Pringle,
  \href{http://adsabs.harvard.edu/cgi-bin/nph-data\_query?bibcode=2011CoPhC.182..946M\&link\_type=EJOURNAL}{{A.I.K.E.F.:
  Adaptive hybrid model for space plasma simulations}}, Computer Physics
  Communications 182~(4) (2011) 946 -- 966.
\newblock \href {https://doi.org/10.1016/j.cpc.2010.12.033}
  {\path{doi:10.1016/j.cpc.2010.12.033}}.
\newline\urlprefix\url{http://adsabs.harvard.edu/cgi-bin/nph-data\_query?bibcode=2011CoPhC.182..946M\&link\_type=EJOURNAL}

\bibitem{fujimoto2011}
K.~Fujimoto,
  \href{http://adsabs.harvard.edu/cgi-bin/nph-data\_query?bibcode=2011JCoPh.230.8508F\&link\_type=EJOURNAL}{{A
  new electromagnetic particle-in-cell model with adaptive mesh refinement for
  high-performance parallel computation}}, Journal of Computational Physics
  230~(2) (2011) 8508 -- 8526.
\newblock \href {https://doi.org/10.1016/j.jcp.2011.08.002}
  {\path{doi:10.1016/j.jcp.2011.08.002}}.
\newline\urlprefix\url{http://adsabs.harvard.edu/cgi-bin/nph-data\_query?bibcode=2011JCoPh.230.8508F\&link\_type=EJOURNAL}

\bibitem{Fujimoto:2018gi}
K.~Fujimoto,
  \href{https://www.frontiersin.org/article/10.3389/fphy.2018.00119/full}{{Multi-Scale
  Kinetic Simulation of Magnetic Reconnection With Dynamically Adaptive
  Meshes}}, Frontiers in Physics 6 (2018) 303 -- 8.
\newblock \href {https://doi.org/10.3389/fphy.2018.00119}
  {\path{doi:10.3389/fphy.2018.00119}}.
\newline\urlprefix\url{https://www.frontiersin.org/article/10.3389/fphy.2018.00119/full}

\bibitem{Lapenta:2007tj}
G.~Lapenta, {Automatic adpative multi-dimensional Particle In Cell}, advanced
  methods for space simulations (2007) 1 -- 16.

\bibitem{lapenta2011democritus}
G.~Lapenta,
  \href{http://adsabs.harvard.edu/cgi-bin/nph-data\_query?bibcode=2011JCoPh.230.4679L\&link\_type=EJOURNAL}{{DEMOCRITUS:
  An adaptive particle in cell (PIC) code for object-plasma interactions}},
  Journal of Computational Physics 230~(1) (2011) 4679 -- 4695.
\newblock \href {https://doi.org/10.1016/j.jcp.2011.02.041}
  {\path{doi:10.1016/j.jcp.2011.02.041}}.
\newline\urlprefix\url{http://adsabs.harvard.edu/cgi-bin/nph-data\_query?bibcode=2011JCoPh.230.4679L\&link\_type=EJOURNAL}

\bibitem{fryxell2008}
B.~Fryxell, K.~Olson, P.~Ricker, F.~X. Timmes, M.~Zingale, D.~Q. Lamb,
  P.~MacNeice, R.~Rosner, J.~W. Truran, H.~Tufo, {FLASH: An Adaptive Mesh
  Hydrodynamics Code for Modeling Astrophysical Thermonuclear Flashes}, The
  Astrophysical Journal Supplement Series 131~(1) (2008) 273.
\newblock \href {https://doi.org/10.1086/317361} {\path{doi:10.1086/317361}}.

\bibitem{2012ApJS..198....7M}
A.~Mignone, C.~Zanni, P.~Tzeferacos, B.~v. Straalen, P.~Colella, G.~Bodo,
  \href{http://adsabs.harvard.edu/cgi-bin/nph-data\_query?bibcode=2012ApJS..198....7M\&link\_type=EJOURNAL}{{The
  PLUTO Code for Adaptive Mesh Computations in Astrophysical Fluid Dynamics}},
  The Astrophysical Journal Supplement 198~(1) (2012) 7.
\newblock \href {https://doi.org/10.1088/0067-0049/198/1/7}
  {\path{doi:10.1088/0067-0049/198/1/7}}.
\newline\urlprefix\url{http://adsabs.harvard.edu/cgi-bin/nph-data\_query?bibcode=2012ApJS..198....7M\&link\_type=EJOURNAL}

\bibitem{2008JCoPh.227.6967T}
G.~Tóth, Y.~Ma, T.~I. Gombosi,
  \href{http://adsabs.harvard.edu/cgi-bin/nph-data\_query?bibcode=2008JCoPh.227.6967T\&link\_type=EJOURNAL}{{Hall
  magnetohydrodynamics on block-adaptive grids}}, Journal of Computational
  Physics 227~(1) (2008) 6967 -- 6984.
\newblock \href {https://doi.org/10.1016/j.jcp.2008.04.010}
  {\path{doi:10.1016/j.jcp.2008.04.010}}.
\newline\urlprefix\url{http://adsabs.harvard.edu/cgi-bin/nph-data\_query?bibcode=2008JCoPh.227.6967T\&link\_type=EJOURNAL}

\bibitem{holst2007}
B.~V.~d. Holst, R.~Keppens,
  \href{http://adsabs.harvard.edu/cgi-bin/nph-data\_query?bibcode=2007JCoPh.226..925V\&link\_type=EJOURNAL}{{Hybrid
  block-AMR in cartesian and curvilinear coordinates: MHD applications}},
  Journal of Computational Physics 226~(1) (2007) 925 -- 946.
\newblock \href {https://doi.org/10.1016/j.jcp.2007.05.007}
  {\path{doi:10.1016/j.jcp.2007.05.007}}.
\newline\urlprefix\url{http://adsabs.harvard.edu/cgi-bin/nph-data\_query?bibcode=2007JCoPh.226..925V\&link\_type=EJOURNAL}

\bibitem{Feyerabend:2015ib}
M.~Feyerabend, S.~Simon, U.~Motschmann, L.~Liuzzo,
  \href{http://dx.doi.org/10.1016/j.pss.2015.07.008}{{Filamented ion tail
  structures at Titan\_ A hybrid simulation study}}, Planetary and Space
  Science 117~(C) (2015) 362 -- 376.
\newblock \href {https://doi.org/10.1016/j.pss.2015.07.008}
  {\path{doi:10.1016/j.pss.2015.07.008}}.
\newline\urlprefix\url{http://dx.doi.org/10.1016/j.pss.2015.07.008}

\bibitem{Vernisse:2017ga}
Y.~Vernisse, J.~A. Riousset, U.~Motschmann, K.~Glassmeier,
  \href{http://dx.doi.org/10.1016/j.pss.2016.08.012}{{Stellar winds and
  planetary bodies simulations\_ Magnetized obstacles in super-Alfvénic and
  sub-Alfvénic flows}}, Planetary and Space Science 137~(C) (2017) 40 -- 51.
\newblock \href {https://doi.org/10.1016/j.pss.2016.08.012}
  {\path{doi:10.1016/j.pss.2016.08.012}}.
\newline\urlprefix\url{http://dx.doi.org/10.1016/j.pss.2016.08.012}

\bibitem{Exner:2018fj}
W.~Exner, D.~Heyner, L.~Liuzzo, U.~Motschmann, D.~Shiota, K.~Kusano,
  T.~Shibayama, \href{https://doi.org/10.1016/j.pss.2017.12.016}{{Coronal mass
  ejection hits mercury: A.I.K.E.F. hybrid-code results compared to MESSENGER
  data}}, Planetary and Space Science 153 (2018) 89 -- 99.
\newblock \href {https://doi.org/10.1016/j.pss.2017.12.016}
  {\path{doi:10.1016/j.pss.2017.12.016}}.
\newline\urlprefix\url{https://doi.org/10.1016/j.pss.2017.12.016}

\bibitem{arnold2020}
H.~Arnold, L.~Liuzzo, S.~Simon,
  \href{http://adsabs.harvard.edu/cgi-bin/nph-data\_query?bibcode=2020JGRA..12527346A\&link\_type=EJOURNAL}{{Plasma
  Interaction Signatures of Plumes at Europa}}, Journal of Geophysical Research
  (Space Physics) 125~(1) (2020) e27346.
\newblock \href {https://doi.org/10.1029/2019ja027346}
  {\path{doi:10.1029/2019ja027346}}.
\newline\urlprefix\url{http://adsabs.harvard.edu/cgi-bin/nph-data\_query?bibcode=2020JGRA..12527346A\&link\_type=EJOURNAL}

\bibitem{lapenta2002rezoning}
G.~Lapenta,
  \href{http://adsabs.harvard.edu/cgi-bin/nph-data\_query?bibcode=2002JCoPh.181..317L\&link\_type=EJOURNAL}{{Particle
  Rezoning for Multidimensional Kinetic Particle-In-Cell Simulations}}, Journal
  of Computational Physics 181~(1) (2002) 317 -- 337.
\newblock \href {https://doi.org/10.1006/jcph.2002.7126}
  {\path{doi:10.1006/jcph.2002.7126}}.
\newline\urlprefix\url{http://adsabs.harvard.edu/cgi-bin/nph-data\_query?bibcode=2002JCoPh.181..317L\&link\_type=EJOURNAL}

\bibitem{innocentiMLMD1}
M.~E. Innocenti, G.~Lapenta, S.~Markidis, A.~Beck, A.~Vapirev,
  \href{http://adsabs.harvard.edu/cgi-bin/nph-data\_query?bibcode=2013JCoPh.238..115I\&link\_type=EJOURNAL}{{A
  Multi Level Multi Domain Method for Particle In Cell plasma simulations}},
  Journal of Computational Physics 238~(3) (2013) 115 -- 140.
\newblock \href {https://doi.org/10.1016/j.jcp.2012.12.028}
  {\path{doi:10.1016/j.jcp.2012.12.028}}.
\newline\urlprefix\url{http://adsabs.harvard.edu/cgi-bin/nph-data\_query?bibcode=2013JCoPh.238..115I\&link\_type=EJOURNAL}

\bibitem{beckMLMD2}
A.~Beck, M.~E. Innocenti, G.~Lapenta, S.~Markidis,
  \href{http://adsabs.harvard.edu/cgi-bin/nph-data\_query?bibcode=2014JCoPh.271..430B\&link\_type=EJOURNAL}{{Multi-level
  multi-domain algorithm implementation for two-dimensional multiscale particle
  in cell simulations}}, Journal of Computational Physics 271 (2014) 430 --
  443.
\newblock \href {https://doi.org/10.1016/j.jcp.2013.12.016}
  {\path{doi:10.1016/j.jcp.2013.12.016}}.
\newline\urlprefix\url{http://adsabs.harvard.edu/cgi-bin/nph-data\_query?bibcode=2014JCoPh.271..430B\&link\_type=EJOURNAL}

\bibitem{innocentiMLMDjp}
M.~E. Innocenti, C.~Tronci, S.~Markidis, G.~Lapenta,
  \href{http://adsabs.harvard.edu/cgi-bin/nph-data\_query?bibcode=2016JPhCS.719a2019I\&link\_type=EJOURNAL}{{Grid
  coupling mechanism in the semi-implicit adaptive Multi-Level Multi-Domain
  method}}, Journal of Physics: Conference Series 719~(1) (2016) 012019.
\newblock \href {https://doi.org/10.1088/1742-6596/719/1/012019}
  {\path{doi:10.1088/1742-6596/719/1/012019}}.
\newline\urlprefix\url{http://adsabs.harvard.edu/cgi-bin/nph-data\_query?bibcode=2016JPhCS.719a2019I\&link\_type=EJOURNAL}

\bibitem{innocentiMLMDmomentum}
M.~E. Innocenti, A.~Beck, S.~Markidis, G.~Lapenta,
  \href{http://adsabs.harvard.edu/cgi-bin/nph-data\_query?bibcode=2016JCoPh.312...14I\&link\_type=EJOURNAL}{{Momentum
  conservation in Multi-Level Multi-Domain (MLMD) simulations}}, Journal of
  Computational Physics 312 (2016) 14 -- 18.
\newblock \href {https://doi.org/10.1016/j.jcp.2016.02.026}
  {\path{doi:10.1016/j.jcp.2016.02.026}}.
\newline\urlprefix\url{http://adsabs.harvard.edu/cgi-bin/nph-data\_query?bibcode=2016JCoPh.312...14I\&link\_type=EJOURNAL}

\bibitem{innocentiMLMDdt}
M.~E. Innocenti, A.~Beck, T.~Ponweiser, S.~Markidis, G.~Lapenta,
  \href{http://adsabs.harvard.edu/cgi-bin/nph-data\_query?bibcode=2015CoPhC.189...47I\&link\_type=EJOURNAL}{{Introduction
  of temporal sub-stepping in the Multi-Level Multi-Domain semi-implicit
  Particle-In-Cell code Parsek2D-MLMD}}, Computer Physics Communications 189
  (2015) 47 -- 59.
\newblock \href {https://doi.org/10.1016/j.cpc.2014.12.004}
  {\path{doi:10.1016/j.cpc.2014.12.004}}.
\newline\urlprefix\url{http://adsabs.harvard.edu/cgi-bin/nph-data\_query?bibcode=2015CoPhC.189...47I\&link\_type=EJOURNAL}

\bibitem{2007JCoPh.227.1340S}
T.~Sugiyama, K.~Kusano,
  \href{http://adsabs.harvard.edu/cgi-bin/nph-data\_query?bibcode=2007JCoPh.227.1340S\&link\_type=EJOURNAL}{{Multi-scale
  plasma simulation by the interlocking of magnetohydrodynamic model and
  particle-in-cell kinetic model}}, Journal of Computational Physics 227~(2)
  (2007) 1340 -- 1352.
\newblock \href {https://doi.org/10.1016/j.jcp.2007.09.011}
  {\path{doi:10.1016/j.jcp.2007.09.011}}.
\newline\urlprefix\url{http://adsabs.harvard.edu/cgi-bin/nph-data\_query?bibcode=2007JCoPh.227.1340S\&link\_type=EJOURNAL}

\bibitem{Ishiguro:2010cu}
S.~Ishiguro, S.~Usami, R.~Horiuchi, H.~Ohtani, A.~Maluckov, M.~M. Škorić,
  \href{https://iopscience.iop.org/article/10.1088/1742-6596/257/1/012026}{{Multi-scale
  simulation for plasma science}}, Journal of Physics: Conference Series
  257~(1) (2010) 012026 -- 9.
\newblock \href {https://doi.org/10.1088/1742-6596/257/1/012026}
  {\path{doi:10.1088/1742-6596/257/1/012026}}.
\newline\urlprefix\url{https://iopscience.iop.org/article/10.1088/1742-6596/257/1/012026}

\bibitem{Usami:2013bc}
S.~Usami, R.~Horiuchi, H.~Ohtani, M.~den,
  \href{http://aip.scitation.org/doi/10.1063/1.4811121}{{Development of
  multi-hierarchy simulation model with non-uniform space grids for
  collisionless driven reconnection}}, Physics of Plasmas 20~(6) (2013) 061208
  -- 9.
\newblock \href {https://doi.org/10.1063/1.4811121}
  {\path{doi:10.1063/1.4811121}}.
\newline\urlprefix\url{http://aip.scitation.org/doi/10.1063/1.4811121}

\bibitem{daldorff2014}
L.~K.~S. Daldorff, G.~Tóth, T.~I. Gombosi, G.~Lapenta, J.~Amaya, S.~Markidis,
  J.~U. Brackbill,
  \href{http://adsabs.harvard.edu/cgi-bin/nph-data\_query?bibcode=2014JCoPh.268..236D\&link\_type=EJOURNAL}{{Two-way
  coupling of a global Hall magnetohydrodynamics model with a local implicit
  particle-in-cell model}}, Journal of Computational Physics 268 (2014) 236 --
  254.
\newblock \href {https://doi.org/10.1016/j.jcp.2014.03.009}
  {\path{doi:10.1016/j.jcp.2014.03.009}}.
\newline\urlprefix\url{http://adsabs.harvard.edu/cgi-bin/nph-data\_query?bibcode=2014JCoPh.268..236D\&link\_type=EJOURNAL}

\bibitem{tothEPIC2016}
G.~Tóth, X.~Jia, S.~Markidis, I.~B. Peng, Y.~Chen, L.~K.~S. Daldorff, V.~M.
  Tenishev, D.~Borovikov, J.~D. Haiducek, T.~I. Gombosi, A.~Glocer, J.~C.
  Dorelli,
  \href{http://adsabs.harvard.edu/cgi-bin/nph-data\_query?bibcode=2016JGRA..121.1273T\&link\_type=EJOURNAL}{{Extended
  magnetohydrodynamics with embedded particle-in-cell simulation of Ganymede's
  magnetosphere}}, Journal of Geophysical Research (Space Physics) 121~(2)
  (2016) 1273 -- 1293.
\newblock \href {https://doi.org/10.1002/2015ja021997}
  {\path{doi:10.1002/2015ja021997}}.
\newline\urlprefix\url{http://adsabs.harvard.edu/cgi-bin/nph-data\_query?bibcode=2016JGRA..121.1273T\&link\_type=EJOURNAL}

\bibitem{chen17MHDEPIC}
Y.~Chen, G.~Tóth, P.~Cassak, X.~Jia, T.~I. Gombosi, J.~A. Slavin, S.~Markidis,
  I.~B. Peng, V.~K. Jordanova, M.~G. Henderson,
  \href{http://adsabs.harvard.edu/cgi-bin/nph-data\_query?bibcode=2017JGRA..12210318C\&link\_type=EJOURNAL}{{Global
  Three-Dimensional Simulation of Earth's Dayside Reconnection Using a Two-Way
  Coupled Magnetohydrodynamics With Embedded Particle-in-Cell Model: Initial
  Results}}, Journal of Geophysical Research (Space Physics) 122~(1) (2017) 10.
\newblock \href {https://doi.org/10.1002/2017ja024186}
  {\path{doi:10.1002/2017ja024186}}.
\newline\urlprefix\url{http://adsabs.harvard.edu/cgi-bin/nph-data\_query?bibcode=2017JGRA..12210318C\&link\_type=EJOURNAL}

\bibitem{zhou19MHDEPIC}
H.~Zhou, G.~Tóth, X.~Jia, Y.~Chen, S.~Markidis,
  \href{http://adsabs.harvard.edu/cgi-bin/nph-data\_query?bibcode=2019JGRA..124.5441Z\&link\_type=EJOURNAL}{{Embedded
  Kinetic Simulation of Ganymede's Magnetosphere: Improvements and
  Inferences}}, Journal of Geophysical Research (Space Physics) 124~(7) (2019)
  5441 -- 5460.
\newblock \href {https://doi.org/10.1029/2019ja026643}
  {\path{doi:10.1029/2019ja026643}}.
\newline\urlprefix\url{http://adsabs.harvard.edu/cgi-bin/nph-data\_query?bibcode=2019JGRA..124.5441Z\&link\_type=EJOURNAL}

\bibitem{Zhou:2020ea}
H.~Zhou, G.~Tóth, X.~Jia, Y.~Chen,
  \href{https://onlinelibrary.wiley.com/doi/10.1029/2020JA028162}{{Reconnection‐Driven
  Dynamics at Ganymede's Upstream Magnetosphere: 3‐D Global Hall MHD and
  MHD‐EPIC Simulations}}, Journal of Geophysical Research (Space Physics)
  125~(8) (2020) 1 -- 20.
\newblock \href {https://doi.org/10.1029/2020ja028162}
  {\path{doi:10.1029/2020ja028162}}.
\newline\urlprefix\url{https://onlinelibrary.wiley.com/doi/10.1029/2020JA028162}

\bibitem{Ma:2018es}
Y.~Ma, C.~T. Russell, G.~Tóth, Y.~Chen, A.~F. Nagy, Y.~Harada, J.~McFadden,
  J.~S. Halekas, R.~Lillis, J.~E.~P. Connerney, J.~Espley, G.~A. DiBraccio,
  S.~Markidis, I.~B. Peng, X.~Fang, B.~M. Jakosky,
  \href{http://doi.wiley.com/10.1029/2017JA024729}{{Reconnection in the Martian
  Magnetotail: Hall-MHD With Embedded Particle-in-Cell Simulations}}, Journal
  of Geophysical Research (Space Physics) 123~(5) (2018) 3742 -- 3763.
\newblock \href {https://doi.org/10.1029/2017ja024729}
  {\path{doi:10.1029/2017ja024729}}.
\newline\urlprefix\url{http://doi.wiley.com/10.1029/2017JA024729}

\bibitem{ChenYuxi2019MHDEPIC}
Y.~Chen, G.~Tóth, X.~Jia, J.~A. Slavin, W.~Sun, S.~Markidis, T.~I. Gombosi,
  J.~M. Raines,
  \href{http://adsabs.harvard.edu/cgi-bin/nph-data\_query?bibcode=2019JGRA..124.8954C\&link\_type=EJOURNAL}{{Studying
  Dawn-Dusk Asymmetries of Mercury's Magnetotail Using MHD-EPIC Simulations}},
  Journal of Geophysical Research (Space Physics) 124~(1) (2019) 8954 -- 8973.
\newblock \href {https://doi.org/10.1029/2019ja026840}
  {\path{doi:10.1029/2019ja026840}}.
\newline\urlprefix\url{http://adsabs.harvard.edu/cgi-bin/nph-data\_query?bibcode=2019JGRA..124.8954C\&link\_type=EJOURNAL}

\bibitem{2017CoPhC.221...81M}
K.~D. Makwana, R.~Keppens, G.~Lapenta,
  \href{http://adsabs.harvard.edu/cgi-bin/nph-data\_query?bibcode=2017CoPhC.221...81M\&link\_type=EJOURNAL}{{Two-way
  coupling of magnetohydrodynamic simulations with embedded particle-in-cell
  simulations}}, Computer Physics Communications 221 (2017) 81 -- 94.
\newblock \href {https://doi.org/10.1016/j.cpc.2017.08.003}
  {\path{doi:10.1016/j.cpc.2017.08.003}}.
\newline\urlprefix\url{http://adsabs.harvard.edu/cgi-bin/nph-data\_query?bibcode=2017CoPhC.221...81M\&link\_type=EJOURNAL}

\bibitem{Makwana:2018jd}
K.~D. Makwana, R.~Keppens, G.~Lapenta,
  \href{http://aip.scitation.org/doi/10.1063/1.5037774}{{Study of magnetic
  reconnection in large-scale magnetic island coalescence via spatially coupled
  MHD and PIC simulations}}, Physics of Plasmas 25~(8) (2018) 082904 -- 12.
\newblock \href {https://doi.org/10.1063/1.5037774}
  {\path{doi:10.1063/1.5037774}}.
\newline\urlprefix\url{http://aip.scitation.org/doi/10.1063/1.5037774}

\bibitem{1967IBMJ...11..215C}
R.~{Courant}, K.~{Friedrichs}, H.~{Lewy}, {On the Partial Difference Equations
  of Mathematical Physics}, IBM Journal of Research and Development 11 (1967)
  215--234.
\newblock \href {https://doi.org/10.1147/rd.112.0215}
  {\path{doi:10.1147/rd.112.0215}}.

\bibitem{aunai2013dissip}
N.~{Aunai}, M.~{Hesse}, C.~{Black}, R.~{Evans}, M.~{Kuznetsova}, {Influence of
  the dissipation mechanism on collisionless magnetic reconnection in symmetric
  and asymmetric current layers}, Physics of Plasmas 20~(4) (2013) 042901.
\newblock \href {http://arxiv.org/abs/1303.0442} {\path{arXiv:1303.0442}},
  \href {https://doi.org/10.1063/1.4795727} {\path{doi:10.1063/1.4795727}}.

\bibitem{Yee}
K.~{Yee}, {Numerical solution of inital boundary value problems involving
  maxwell's equations in isotropic media}, IEEE Transactions on Antennas and
  Propagation 14~(3) (1966) 302--307.
\newblock \href {https://doi.org/10.1109/TAP.1966.1138693}
  {\path{doi:10.1109/TAP.1966.1138693}}.

\bibitem{kunz}
M.~W. {Kunz}, J.~M. {Stone}, X.-N. {Bai}, {Pegasus: A new hybrid-kinetic
  particle-in-cell code for astrophysical plasma dynamics}, Journal of
  Computational Physics 259 (2014) 154--174.
\newblock \href {http://arxiv.org/abs/1311.4865} {\path{arXiv:1311.4865}},
  \href {https://doi.org/10.1016/j.jcp.2013.11.035}
  {\path{doi:10.1016/j.jcp.2013.11.035}}.

\bibitem{Boris1970}
J.~P. Boris, \href{http://scholar.google.comjavascript:void(0)}{{Fourth
  Conference on Numerical Simulation of Plasmas}}, 1970.
\newline\urlprefix\url{http://scholar.google.comjavascript:void(0)}

\bibitem{lohner87}
R.~Löhner, {An adaptive finite element scheme for transient problems in CFD},
  Computer Methods in Applied Mechanics and Engineering 61~(3) (1987) 323--338.
\newblock \href {https://doi.org/10.1016/0045-7825(87)90098-3}
  {\path{doi:10.1016/0045-7825(87)90098-3}}.

\bibitem{gunney16}
B.~T.~N. Gunney, R.~W. Anderson, {Advances in patch-based adaptive mesh
  refinement scalability}, Journal of Parallel and Distributed Computing 89
  (2016) 65--84.
\newblock \href {https://doi.org/10.1016/j.jpdc.2015.11.005}
  {\path{doi:10.1016/j.jpdc.2015.11.005}}.

\bibitem{gary1993}
S.~P. {Gary}, Theory of Space Plasma Microinstabilities, 1993.

\bibitem{gary1985}
S.~P. {Gary}, {Electromagnetic ion beam instabilities - Hot beams at
  interplanetary shocks}, Astrophysical Journal 288 (1985) 342--352.
\newblock \href {https://doi.org/10.1086/162797} {\path{doi:10.1086/162797}}.

\bibitem{1998JGR...103.9165S}
M.~Shay, J.~Drake, R.~Denton, D.~Biskamp,
  \href{http://adsabs.harvard.edu/cgi-bin/nph-data\_query?bibcode=1998JGR...103.9165S\&link\_type=ABSTRACT}{{Structure
  of the dissipation region during collisionless magnetic reconnection}},
  Journal of Geophysical Research 103~(A) (1998) 9165 -- 9176.
\newblock \href {https://doi.org/10.1029/97ja03528}
  {\path{doi:10.1029/97ja03528}}.
\newline\urlprefix\url{http://adsabs.harvard.edu/cgi-bin/nph-data\_query?bibcode=1998JGR...103.9165S\&link\_type=ABSTRACT}

\bibitem{2002APS..DFD.KA002W}
S.~{Wijesinghe}, R.~{Hornung}, A.~{Garcia}, N.~{Hadjiconstantinou}, {Three
  Dimensional Hybrid Continuum-Atomistic Simulations for Multiscale
  Hydrodynamics}, in: APS Division of Fluid Dynamics Meeting Abstracts, Vol.~55
  of APS Meeting Abstracts, 2002, p. KA.002.

\end{thebibliography}

\end{document}